\documentclass[11pt]{svjour3}

\usepackage[utf8]{inputenc}
\usepackage[english]{babel}
\usepackage{amsmath}
\usepackage{amsfonts}
\usepackage{amssymb}
\usepackage{graphicx}
\usepackage{mathrsfs}
\usepackage[left=2cm,right=2cm,top=2cm,bottom=2cm]{geometry}
\usepackage{epstopdf}
\usepackage{pgf}
\usepackage{stmaryrd}
\usepackage[verbose]{wrapfig}
\usepackage{xcolor}
\usepackage{url}
\usepackage{authblk}
\usepackage{graphicx}
\usepackage[verbose]{wrapfig}
\usepackage[margin=1cm]{caption}
\usepackage{subcaption}
\usepackage[]{algorithm2e}
\usepackage{braket}
\usepackage{stackengine}
\usepackage[numbers]{natbib}
\usepackage{dsfont}
\usepackage{color}
\usepackage{framed}

\title{Combinatorial Optimization on Gate Model Quantum Computers: A Survey}
 
\author{Ehsan Zahedinejad \and Arman Zaribafiyan}

\institute{E. Zahedinejad \and A. Zaribafiyan
\at 1QB Information Technologies (1QBit), 458--550 Burrard Street, Vancouver, BC, Canada, \, V6C 2B5
\and
A. Zaribafiyan
\at Department of Electrical and Computer Engineering, University of British Columbia, 2332 Main Mall, Vancouver, BC, Canada \, V6T 1Z4}

\begin{document}
\maketitle
\abstract{The advent of quantum computing processors with possibility to scale beyond experimental capacities magnifies the importance of studying their applications. Combinatorial optimization problems can be one of the promising applications of these new devices.  These problems are recurrent in industrial applications and they are in general difficult for classical computing hardware.
In this work, we provide a survey of the approaches to solving different types of combinatorial optimization problems, in particular quadratic unconstrained binary optimization (QUBO) problems on a gate model quantum computer. 
We  focus mainly on four different approaches including digitizing the adiabatic quantum computing, global quantum optimization algorithms, the quantum algorithms that approximate the ground state of a general QUBO problem, and quantum sampling. We also discuss the quantum algorithms  that are custom designed to solve certain types of QUBO problems.}

\section{Introduction}
Quantum computers~\cite{LJL+10,Llo96} are theoretically proven to solve certain families of computational problems faster than the best known classical algorithms~\cite{LWG+17,Sho99,Gro96}. There are different paradigms to realize quantum computing~\cite{NC00} including (but not limited to)  adiabatic quantum computing (AQC)~\cite{FGG+00} and gate model quantum computing~\cite{Sho96}.\footnote{The gate model paradigm is interchangeably also called the standard approach or the circuit-based approach in the literature.} Each paradigm can be realized within various systems~\cite{JGC+08,CW08,Li809,JAG+11}. Over the past years there has been an impressive progress in the field of AQC~\cite{JAG+11,CFP01,AM06}, with many practical problems within the field of combinatorial optimization are proposed to be solved using this approach~\cite{HZA+16,ZMR+16,VMR15,MW13,RVO+15,HRI15}. As AQC (at zero temperature) and gate model quantum computing are shown to be polynomially equivalent~\cite{ADK+08}, a valid question would be how to solve combinatorial optimization problems on a gate model quantum computer. Here we will address this question by giving a short survey on quantum algorithms for solving combinatorial optimization problems. We mainly focus on quadratic unconstrained binary optimization (QUBO) problems although the arguments can be generalized to any unconstrained discrete variable optimization problem.

The primary objective in quantum algorithm design is to devise algorithms which require less time and resources to find an optimal (or near-optimal) solution to a specific computational task~\cite{KJL+08,KSV02}. Shor's factorization algorithm~\cite{Sho99} is an example of a quantum algorithm that performs prime factorization in polynomial time. Due to Shor's algorithm, the prime factorization is now in the complexity class BQP (bounded-error quantum polynomial time), whereas all of the other existing classical algorithms solve the problem in sub-exponential time~\cite{Len92,Pom82}. The exponential speed-up of Shor's factorization algorithm gives rise to the question of whether other problems in the NP can be reduced to BQP on a quantum computer.

Quadratic unconstrained binary optimization is important in that it provides us with the ability to formulate many applied problems in combinatorial optimization and is defined by:
\begin{align}
\label{eq:optim}
\min \:\: & {\bf x^t Q x} \\ \nonumber
\text{s.t.} \:\: & {\bf x} \in \{0, 1\}^n ,
\end{align}
where $Q$ is an $n$-by-$n$ square, symmetric matrix of real-valued coefficients. Many problems, including graph optimization, scheduling and resource allocation, and clustering and partitioning, can be formulated as a QUBO problem. These problems (and QUBO problems in general) are proved to be NP-complete~\cite{LR73,GJS76,GJS76_2}. The outstanding advantage of formulating these problems as QUBO problems is that often general QUBO solvers, which are not specialized to exploit problem heuristics, provide solutions which are either superior to, or as good as, the best specialized approaches, both in terms of quality and efficiency~\cite{Kochenberger2014}.These reasons, along with the wide range of applications of the mentioned industry problems, make a strong case for studying the feasibility and performance of a general quantum QUBO solver.

\begin{figure}
\centering
\includegraphics[width=.65\textwidth]{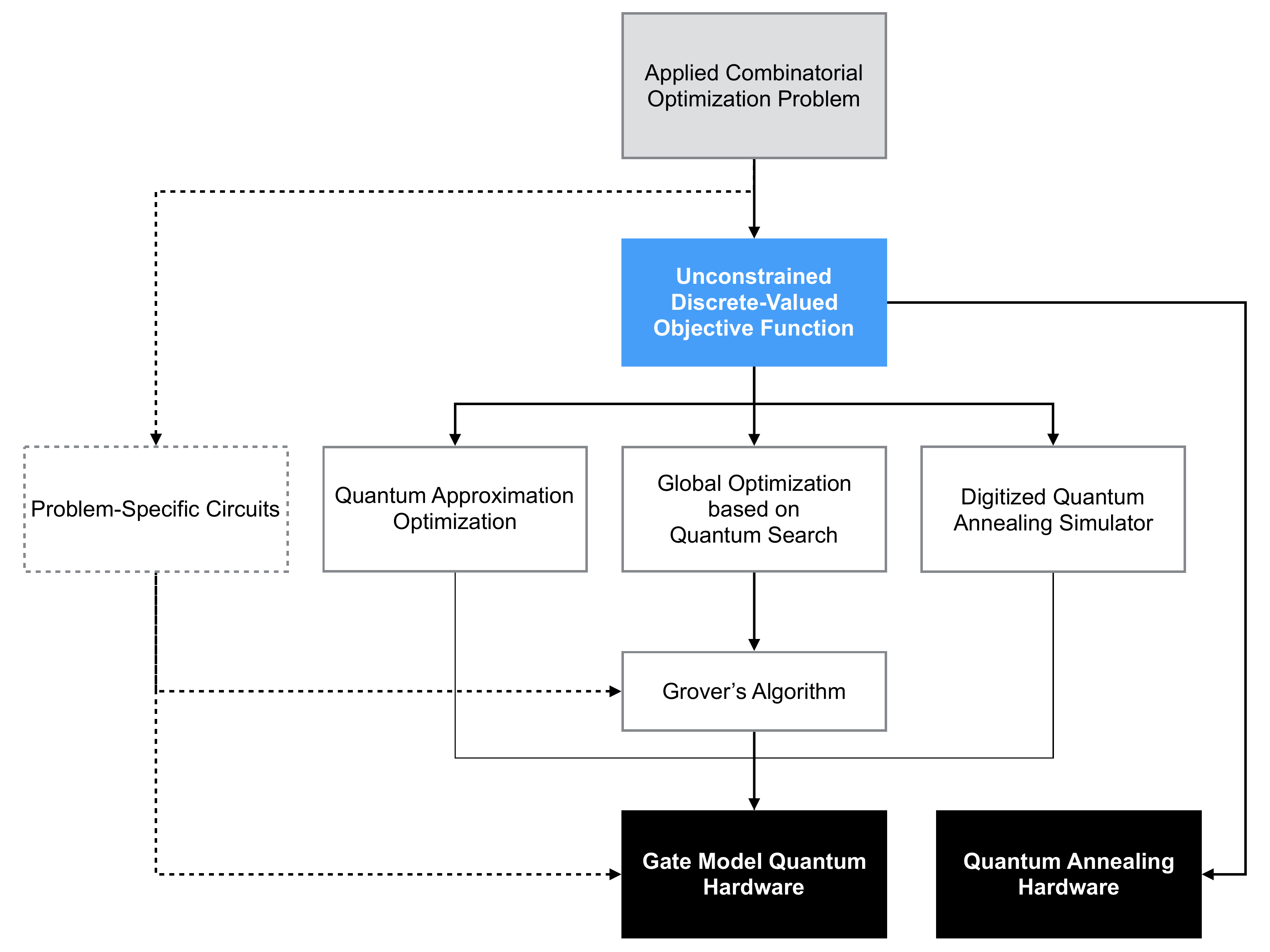}
\caption{A flowchart illustrating how different approaches to solving combinatorial optimization problems can utilize different realizations of quantum computation devices and quantum algorithms.}
\label{fig:diagram}
\end{figure}

Despite many attempts to investigate and improve general quantum (or classical) QUBO solvers \cite{ZCH16,EP02,bian2010ising}, a quantum algorithm for solving general QUBO models in polynomial time has not been found. The optimality of Grover's search algorithm~\cite{Zal99}, which proposes at most a quadratic speed-up for searching an unstructured database~\cite{Gro96}, makes it difficult to find other quantum algorithms for solving general QUBO models. There have been many attempts to investigate the effectiveness of quantum annealers for solving QUBO problems. The results indicate that these devices might also have the potential to solve families of QUBO problems more efficiently than classical computers~\cite{VMK+15,KHZ+15,DBI+16,KYRO+17}. Fig.~\ref{fig:diagram} illustrates the solving of applied combinatorial optimization problems by first formulating them as an unconstrained discrete-valued optimization task (like a QUBO problem) and then using different quantum computation algorithms, devices, and platforms to solve the problem.

Given the fact that there is a large body of literature on solving QUBO problems using adiabatic quantum computation (AQC), we focus in this survey mainly on quantum algorithms that solve QUBO problems on a gate model quantum computer. In particular, we review four distinct approaches to tackling this problem, including the simulation of AQC on a gate model quantum computer~\cite{ADK+08,DMV01}, (the applicability of) Grover's search algorithm for solving problems in the NP complexity class in a global optimization framework~\cite{BBW05}, quantum approximate optimization algorithms (QAOA) which are inspired by AQC~\cite{FGG14}, and, lastly, quantum Metropolis sampling~\cite{TOV+11}.

This survey is by no means a comprehensive review of all of the research and recent work related to the solving of combinatorial optimization problems using quantum algorithms. Rather, we focus on the quantum algorithms that can be implemented on near-term quantum devices, or those quantum algorithms that have been used by other researchers to solve optimization problems. For completeness, whenever possible, we refer readers to the appropriate reference materials that discuss other quantum algorithms relevant to our survey.

This survey is structured as follows. In Section~\ref{sec:QUBO}, we define the Ising model and QUBO problems. In Section~\ref{sec:AQCQUBO}, we explain AQC and its relation to QUBO. In Section~\ref{sec:AQCsim}, we explain the standard method of simulating AQC on a gate model quantum computer. Section~\ref{sec:GQO} focuses on global quantum optimization algorithms, including Grover-based algorithms as well as heuristic approaches. Section~\ref{sec:approxi} discusses the methods that approximate the ground state of the classical Ising model. We discuss quantum Metropolis sampling in Section~\ref{sec:qms}. A review of the quantum algorithms that have been devised to solve specific types of QUBO problems is given in Section~\ref{sec:specific}. We conclude this survey with a brief discussion and propose some future research directions in Section \ref{sec:conclude}.

\section{The Ising model and QUBO problems}
\label{sec:QUBO}
In this section, we first present the mathematical formulation of an Ising problem which is one of the most widely used models in physics~\cite{Cip87}. We then establish the relation between an Ising model and a QUBO formulation. 

An Ising model, first proposed by Ernst Ising and Wilhelm Lenz, explains the interaction of molecules in a magnetic material in an external magnetic field. In a mathematical abstract form, an Ising model consists of a set of spins $\mathcal{V}$, each taking a value of  $s_i\in\{-1,1\}$. Denoting $\mathscr{E}$ as a set of pair-wise interactions between spins, one can formulate the energy $E(\bf{s})$ of the spin system for a configuration $\bf{s}\in\{-1,1\}^\mathcal{|V|}$ using an Ising Hamiltonian:
\begin{equation}
\label{eq:ising}
E({\bf{s}})=\sum_{(i,j)\in\mathscr{E}}J_{i,j}s_is_j+\sum_{i\in\mathcal{V}}h_is_i \, .
\end{equation}

 The Ising model has the QUBO formulation
  \begin{equation}
 \label{eq:qubo}
 f(x_1,\dots,x_n)=c_0+\displaystyle\sum_{j=1}^{n} c_jx_j+\displaystyle\sum_{1\le{i}<{j}\le{n}}^{n} d_{ij}x_ix_j \,, 
 \end{equation}
which is equivalent to (\ref{eq:ising}) up to a linear transformation\footnote{We refer the interested reader to~\cite{bian2010ising} for he explicit steps of the transformation.} ($s_i = 2x_i-1$), where in Equation (\ref{eq:qubo}), $x_i\in\{0,1\}$ and $c_i$ and $d_{ij}$ are real numbers.
Given this mapping between (\ref{eq:ising}) and (\ref{eq:qubo}), solving a QUBO problem is equivalent to finding the ground state of an Ising model Hamiltonian; one can, therefore, utilize quantum algorithms to do so. In the next section, we begin our survey, explaining how AQC is related to the solving of QUBO problems, and how  AQC can be digitized and simulated on a gate model quantum computer to solve such problems.

\section{AQC and QUBO problems}
\label{sec:AQCQUBO}  
\begin{figure}
\centering
\includegraphics[width=.5\textwidth]{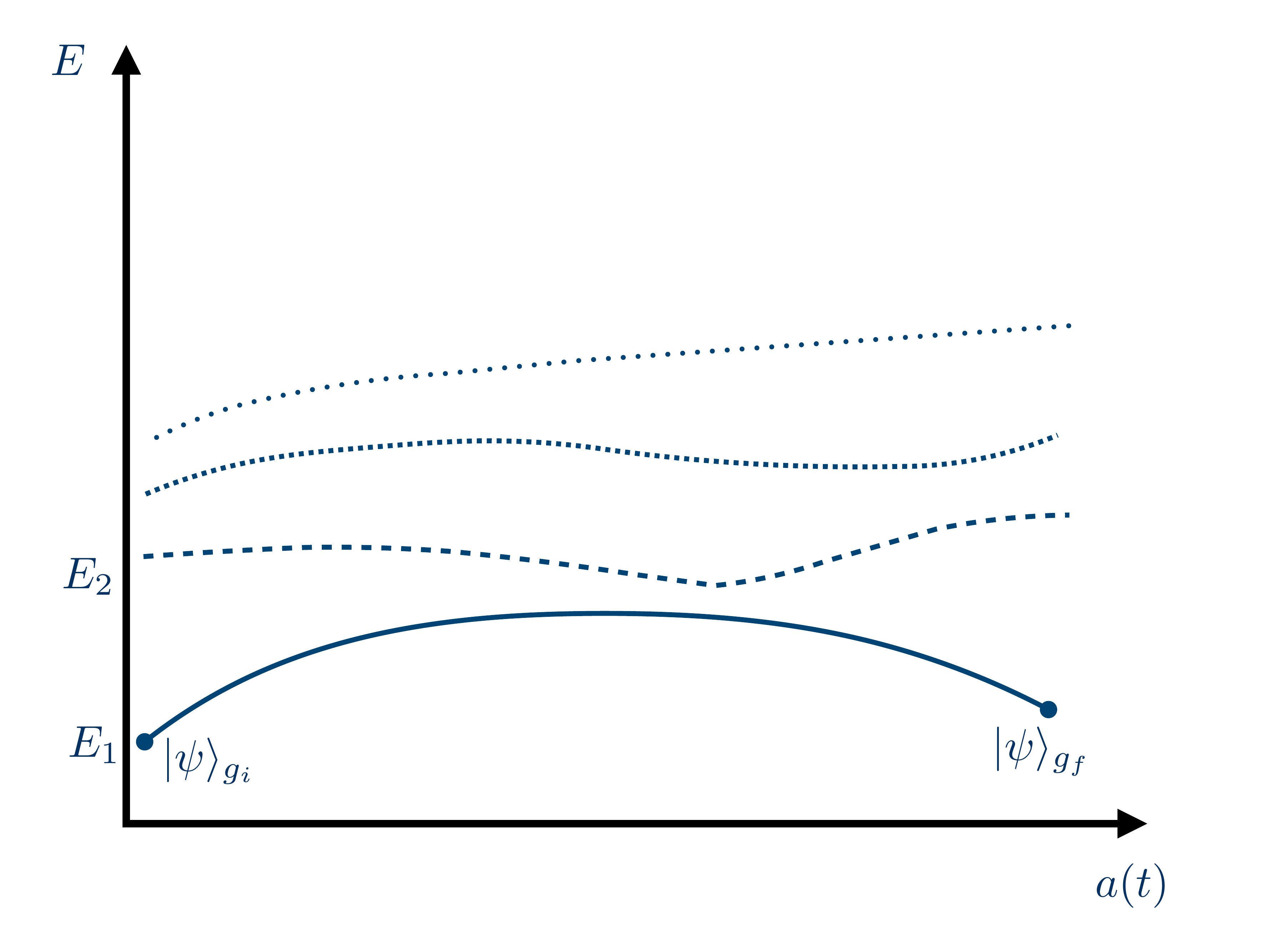}
\caption{A schematic view of the AQC process. The vertical axis denotes the energy spectrum, $\text{E}$, of the whole system Hamiltonian. The horizontal axis shows the external parameter, $a(t)$, changing adiabatically over time. The system is initialized at the ground state $\ket{\psi}_{g_\text{i}}$ of an initial Hamiltonian and then slowly evolved toward the final Hamiltonian with the ground state $\ket{\psi}_{g_\text{f}}$.}
\label{fig:AQC}
\end{figure}

The idea of adiabatic quantum computation was first proposed by Farhi et al.~\cite{FGG+00,FGG+01,farhi2002quantum,farhi2002quantum2}, wherein the goal was to employ a physical quantum computer to solve  a combinatorial optimization problem. Over the past decade, there has been  a great deal of progress in designing adiabatic quantum devices, with the D-Wave 2000Q quantum computing machine, with more than two thousand qubits, being the latest quantum adiabatic optimizer~\cite{TK16}. A pictorial explanation of AQC is given in Fig.~\ref{fig:AQC} . The main idea in AQC is to initialize the system in the known ground state $\ket{\psi}_{g_\text{i}}$ of a specific Hamiltonian. Then, the system Hamiltonian is slowly evolved from the initial Hamiltonian toward the final Hamiltonian in which we have encoded the problem Hamiltonian (equivalently, the QUBO problem) in the form of an Ising model (see (\ref{eq:ising})). The adiabatic theorem states that if the evolution is sufficiently slow, the system will remain in its ground state (or one of its ground states) throughout the entire evolution process. Therefore, starting from the ground state of $H_\text{i}$, the system ends up in the ground state of $H_\text{f}$ (i.e., $\ket{\psi}_{g_\text{f}}$), that is, the solution of the optimization problem in (\ref{eq:optim}). In the next section, we explain how AQC can be simulated on a gate model quantum computer.

\section{Simulating AQC on a gate model quantum computer}
\label{sec:AQCsim}  
It is natural to ask why we would need to simulate AQC on a gate model quantum computer. The answer is that, in practice, we can always use quantum annealing to solve a general QUBO problem. One of the main challenges in designing a universal adiabatic quantum computer is in realizing physical quantum systems that can accommodate any type of non-stoquastic Hamiltonian. The state-of-the-art quantum annealers do not currently provide such interactions. This challenge can be (at least in principle) addressed by decomposing the terms in a generic Hamiltonian into a sequence of single- and two-qubit gates. There exist additional, general challenges to address, such as supporting graphs with higher connectivity and various multi-body interactions, increasing robustness against  noise, and attaining high precision in programming the problem Hamiltonian that can affect the design of a universal adiabatic or gate model quantum computer. It remains to be seen how each of these challenges can be addressed in the design of a universal quantum computing device.

The idea behind digitizing AQC is to decompose the unitary evolution of a quantum system into a sequence of single- and two-qubit gates. There are several approaches to performing this decomposition. In fact, the whole scheme of digitization belongs to a broader field of quantum mechanics called ``quantum Hamiltonian simulation''. In this section, we explain only one of the most straightforward methods of Hamiltonian simulation that has practically been used to simulate AQC on a gate model quantum processor.

In an AQC setting, the system Hamiltonian can be represented by a combination of an initial Hamiltonian ($H_\text{i}$) whose ground state we know and final (problem) Hamiltonian ($H_\text{f}$) whose ground state we aim to find. We denote this combination as
\begin{equation}
\label{eq:AQC1}
{\hat{H}}(t)=(1-a(t))H_\text{i}+a(t)H_\text{f}\,,
\end{equation}
where $\hat{H}$ is the total Hamiltonian of the system and $a(t)$ is a time-dependent function that varies from $a(t=0)=0$ to $a(t=T)=1$, where $T$ is the total evolution time of the system. In order to ease the digitization formulation, we reformulate (\ref{eq:AQC1}) into
\begin{equation}
\label{AQCH}
\hat{H}(t)=\sum_{i=1}^{I}{{a}(t)H_i}\,,
\end{equation}
where $a(t)$ are time-dependent real variables and $H_i$ are $k$-local Hamiltonians, that is, they act on at most $k$ qubits. 
 A simple method to digitize the time evolution of a system with Hamiltonian~(\ref{AQCH}) is to use the first-order Lie--Trotter--Suzuki formula~\cite{Tro59,Suz90}. Using this method, we can approximate the time-ordered ($\mathcal{T}$) evolution operator of the system

\begin{equation}
\label{eq:Ue}
U(T)=\mathcal{T}\text{exp}\left(-i\int_{0}^{T}\hat{H}(t)dt\right)
\end{equation}

\noindent over a time interval [$0,T$] by the approximated operator
\begin{equation}
\label{eq:digitized}
\tilde{U}=\prod_{m=1}^M\prod_{l=1}^I{\text{exp}[-i\delta{t}a_i(m\delta{t})H_i]}\,,
\end{equation}
where, in (\ref{eq:digitized}), $\delta{t}=T/M$ are equally-sized time steps and $M$ is the number of time steps. Given the ability to perform arbitrary local rotations, each $k$-local term in the Hamiltonian (2-local in the case of QUBO) requires at most $\mathcal{O}(k)$ gates. Therefore, in summary, using the above digitization approach, we can employ on the order of $\mathcal{O}(MIk)$ single- or two-qubit logical gates to evolve the initial state of the quantum system (represented as a uniform superposition of all eigenstates of $H_\text{f}$) from
\begin{equation}
\label{eq:inistate}
\ket{\psi}_\text{i}=\frac{1}{\sqrt{2^n}}\sum_{i}\ket{i}
\end{equation}
to a state that represents the ground state of the problem Hamiltonian (i.e., an Ising spin model Hamiltonian with $n$ spins). The preceding argument holds if we know the exact evolution time for the adiabatic process. As we will not be discussing the complexity of finding the ground state of an Ising Hamiltonian, we refer interested readers to~\cite{Bar82} and the references therein.

A similar approach was recently used to digitize AQC on a quantum processor with nine qubits and a thousand quantum logical gates~\cite{BSL+16}. There also exist other methods, in theory, to simulate AQC on a gate model quantum computer with far better complexity over the number of gates required. For example, a method based on Taylor series truncation and approximating the unitary evolution of the system in terms of the sum of some unitary operators is shown to exponentially outperform Trotterization with respect to the  desired precision~\cite{BCC+15}. Currently, the choice of Hamiltonian simulation is largely limited by the size of existing quantum processors, and not by the performance of these methods~\cite{BCK15,BCC+14}.

\section{Global quantum optimization}
\label{sec:GQO}
Most global quantum optimization algorithms employ Grover's search algorithm as a subroutine, and enhance the optimization process by taking advantage of its quadratic speed-up~\cite{BBW05,BBH+98,Liu2010620}. In the first part of this section, we discuss Grover-based optimization algorithms. In the second part, we explain another approach in quantum global optimization which is inspired by AQC.

\subsection{Grover-based quantum global optimization}
\label{sec:Grover}
Grover's search algorithm has been shown to be optimal for searching an unstructured database~\cite{Zal99}, with a quadratic speed-up over the best classical algorithm. As noted by Grover, if the algorithm could solve the unsorted problem in polynomial time on a quantum computer, then it would provide an algorithm for BQP (bounded error quantum polynomial time) for problems in NP~\cite{Gro96}. However, quadratic speed-up of Grover's algorithm and, in fact, its optimal complexity (i.e., $\mathcal{O}(\sqrt{2^n})$, with $n$ being the number of elements in the search database) prohibit any definitive statement about the relation between classes BQP and NP~\cite{Gro96}.

Grover's search algorithm has been widely used in the context of general quantum optimization algorithms. Dur~et~al. first introduced a quantum algorithm for finding the minimum of an unstructured database~\cite{DH96}. Inspired by that work, other researchers also proposed quantum algorithms for optimization problems. Here, we review common fundamental steps of these methods, and explain how they can be used to solve QUBO problems on a gate model quantum computer. We start by defining the problem that Grover's algorithm solves. We then give a geometrical picture of Grover's algorithm~\cite{Mer07} and construct the related quantum circuit. The geometrical picture is then used to explain the effectiveness of Grover-based optimization algorithms. Readers can consult with Grover's original paper for more details~\cite{Gro96}.

Let us define the problem that Grover's algorithm solves. We are given an oracle $\mathcal{Q}$ that, given a question $x$, provides an answer $\mathcal{Q}(x)$, where
\begin{equation}
\label{eq:oracle}
    \mathcal{Q}(x)= 
\begin{cases}
    1,& \text{if } x=c\\
    0,              & \text{otherwise}
\end{cases}
\end{equation}
and $x$ is a $n$-bit integer. We are asked to find the variable $x=c$ using the fewest queries to the oracle. Classically, one can attempt to find the solution, $c$, by making $\mathcal{O}(N)$ queries to the system, where $N=2^n$. Grover's algorithm promises a quadratic speed-up over the best classical algorithm.

The basic idea of Grover's algorithm is to repeatedly apply two operators $\bf{V}$ and $\bf{D}$ (see details about these operators below) to the quantum state
\begin{equation}
\label{eq:spstate}
\ket{\psi}=\frac{1}{\sqrt{2^n}}\sum_{x=0}^{2^n-1}\ket{x}
\end{equation}
such that, through the evolution, the quantum state rotates toward the target solution $\ket{c}$. Here, we assume that the target subspace is unique and there is only one solution $x=c$ that matches the oracle $\mathcal{Q}$. Later, we explain the case where the target subspace includes multiple target states.

The action of operator $\bf{V}$ on (\ref{eq:spstate}) is to flip the sign of the component along the target state $\ket{c}$ while leaving the sign of the other components orthogonal to $\ket{c}$ unchanged. Given the action of $\bf{V}$, one can define the corresponding operator as follows:
\begin{equation}
\label{eq:Vop}
{\bf V}=\mathds{1}-2\ket{c}\bra{c}\,,
\end{equation}
where $\mathds{1}$ is the identity operator and the operator $\bf{D}$, called the diffusion operator, is defined as follows:
\begin{equation}
\label{eq:Dop}
{\bf D}=2\ket{\psi}\bra{\psi}-\mathds{1}.
\end{equation}

\begin{figure}
\centering
\begin{subfigure}{.5\textwidth}
  \centering
  \caption{ }
  \includegraphics[trim ={9.5cm 5.5cm 9.5cm 5.5cm},clip,width=.7\textwidth]{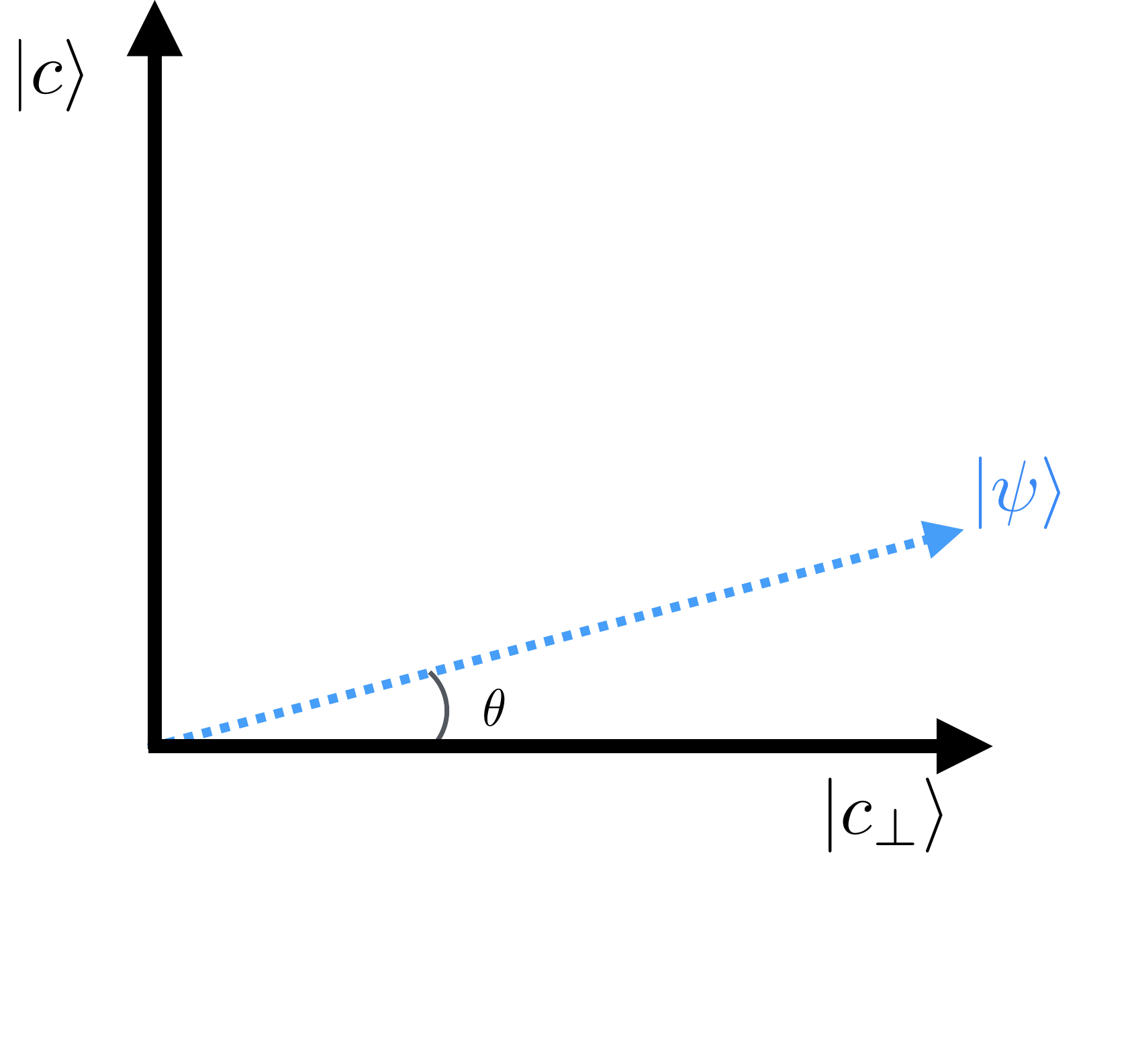}
  
  \label{fig:sub1}
\end{subfigure}%
\begin{subfigure}{.5\textwidth}
  \centering
  \caption{ }
  \includegraphics[trim ={9.5cm 5.5cm 9.5cm 5.5cm},clip, width=.7\textwidth]{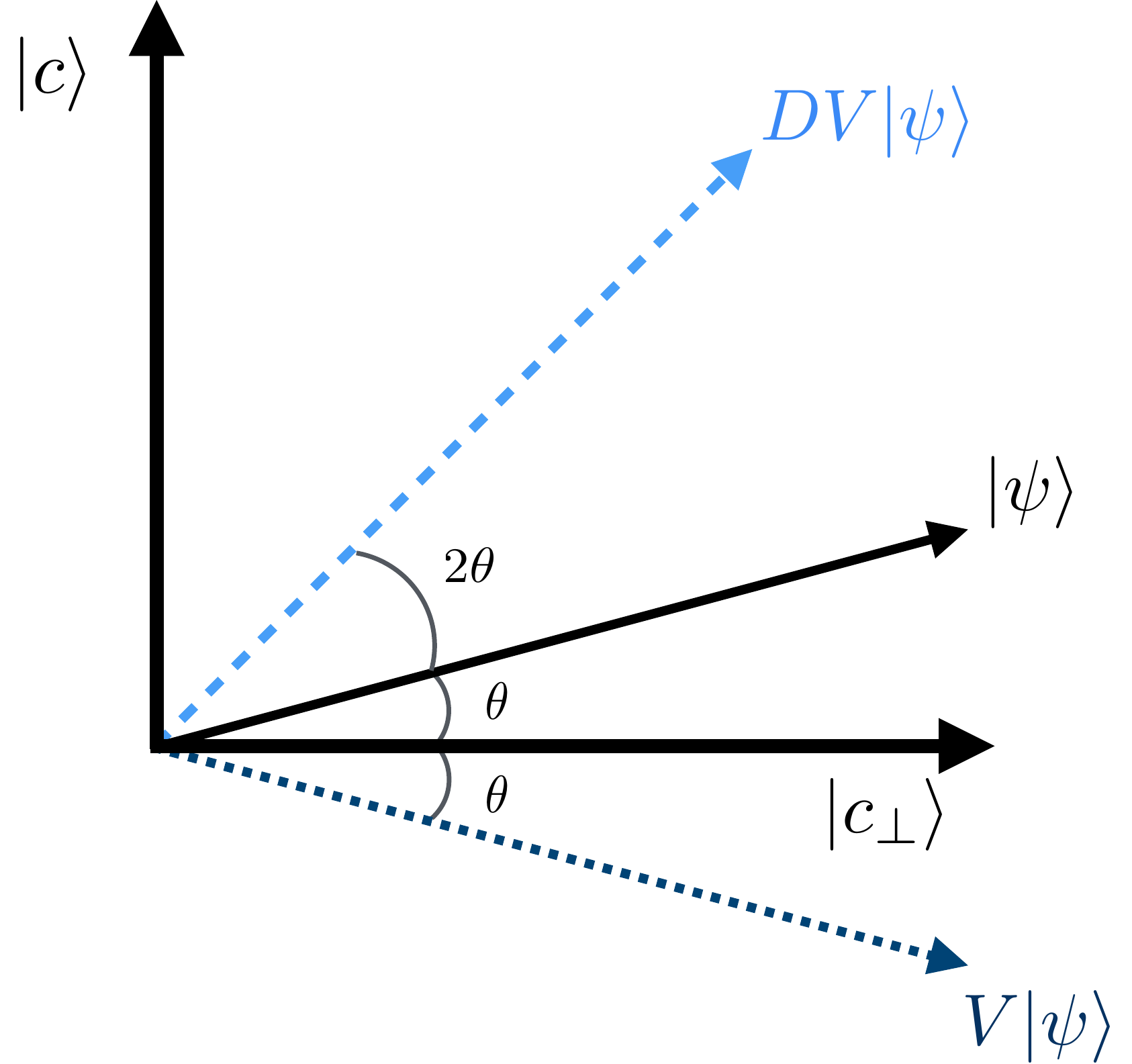}
  
  \label{fig:sub2}
\end{subfigure}
\caption{Geometrical representation of Grover's search algorithm in two-dimensional space. (a) Representing the initial state in the two-dimensional space of the target ($\ket{c}$) and non-target ($\ket{c_\perp}$) states. (b) Action of operators $\bf{D}$ and $\bf{V}$ on the initial state. Each action of $\bf{DV}$ moves the initial state closer to the target state $\ket{c}$ by $2\theta$.}
\label{fig:diagram2}
\end{figure}

As is evident from (\ref{eq:Dop}), the operator $\bf{D}$ preserves the component along the state $\ket{\psi}$ and flips the sign of any state orthogonal to $\ket{\psi}$. It is easy to realize  from (\ref{eq:spstate}) that $\ket{\psi}$ can be represented in the two-dimensional space of $\ket{c}$ (target) and $\ket{c_\perp}$ (non-target) subspace (see Fig.~\ref{fig:diagram2}A), where $\theta$ and $\alpha$ are the angles that $\ket{\psi}$ makes with $\ket{c}$ and $\ket{c_\perp}$, respectively, and where, for large $N$, we have $\theta\approx{2^{-n/2}}$.

The geometrical action of $\bf{V}$ on $\ket{\psi}$ in the two-dimensional plane of $\ket{c}$ and $\ket{c_\perp}$ is to reflect the vector with respect to $\ket{c_\perp}$ as the mirror axis. On the other hand, $\bf{D}$ will reflect the result of the first reflection caused by the action of $\bf{V}$ with respect to $\ket{\psi}$ being the mirror axis of reflection (see Fig.~\ref{fig:diagram2}B). After each operation of $\bf{DV}$, the initial vector $\ket{\psi}$ will rotate toward the target solution by $2\theta$. From this analysis, one can conclude that the total number of rotations  $R$(i.e., the number of times we apply the operations $\bf{DV}$) to the initial state is on the order of
\begin{equation}
\label{eq:rot}
R=\mathcal{O}(2^{n/2}).
\end{equation}

\begin{figure}
\centering
\includegraphics[width=.50\textwidth]{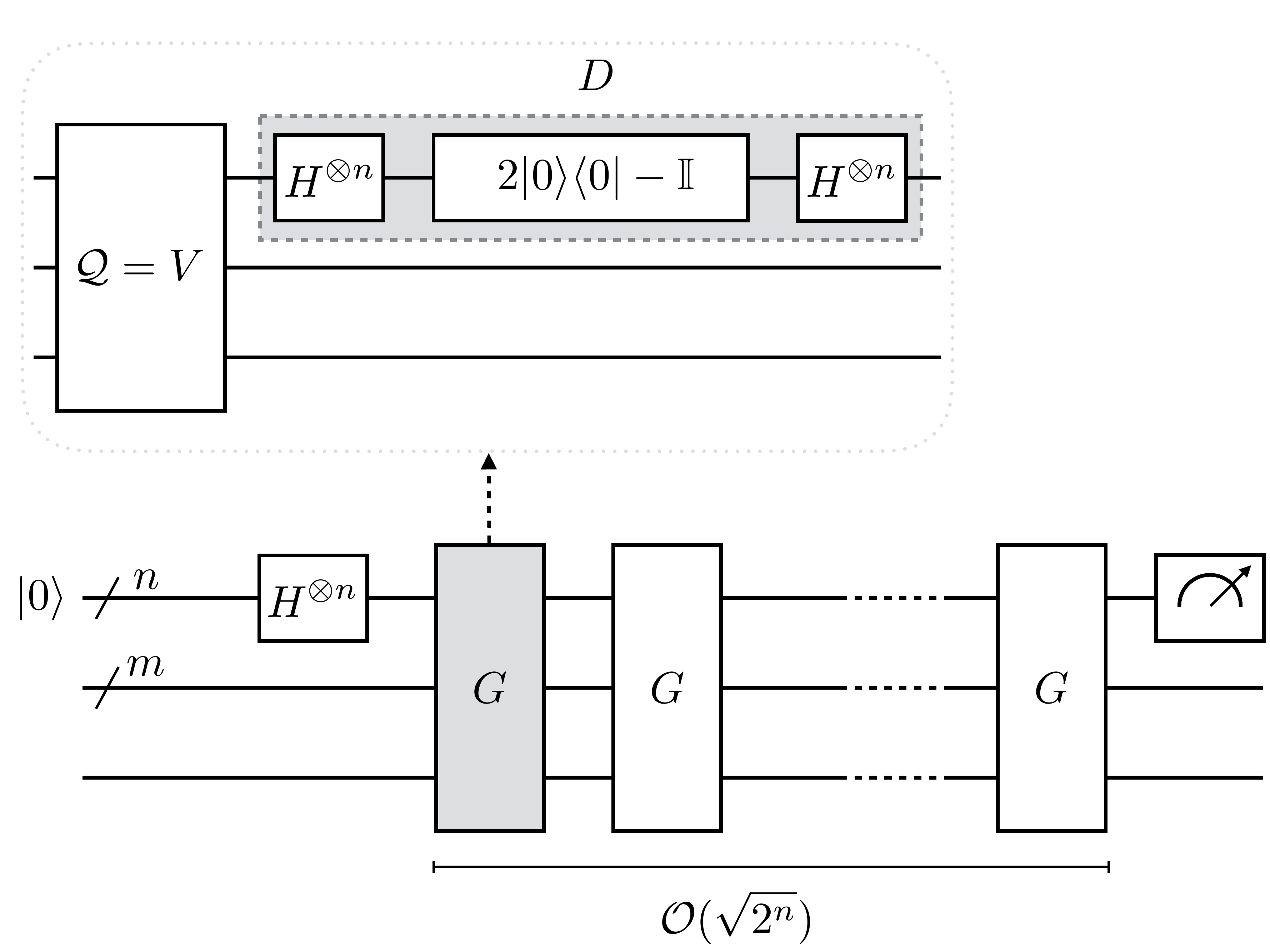}
\caption{Quantum circuit representation of Grover's search algorithm. Each block $G$ in the main circuit (lower part) denotes a Grover search operation. $n$ and $m$ refer to the number of qubits in each register. $H$ is the single-qubit Hadamard gate. $\textbf{V}$ and $\textbf{D}$ refer to the operators discussed in Fig.~\ref{fig:diagram2}. $\mathcal{Q}$ represents the oracle defined in (\ref{eq:oracle}).}
\label{fig:grovercircuit}
\end{figure}

The role of the two operators $\textbf{V}$ and $\textbf{D}$ will be more clear when we show their actions within the quantum circuit of Grover's search algorithm, depicted in Fig.~\ref{fig:grovercircuit}. As is evident from this figure, the operator $\textbf{V}$ implements the oracle $\mathcal{Q}$, whereas $\textbf{D}$ performs a diffusion operation on the problem register with $n$ qubits.
 
\begin{figure}
\centering
\includegraphics[trim ={6cm 0 6cm 0},clip, width=.40\textwidth]{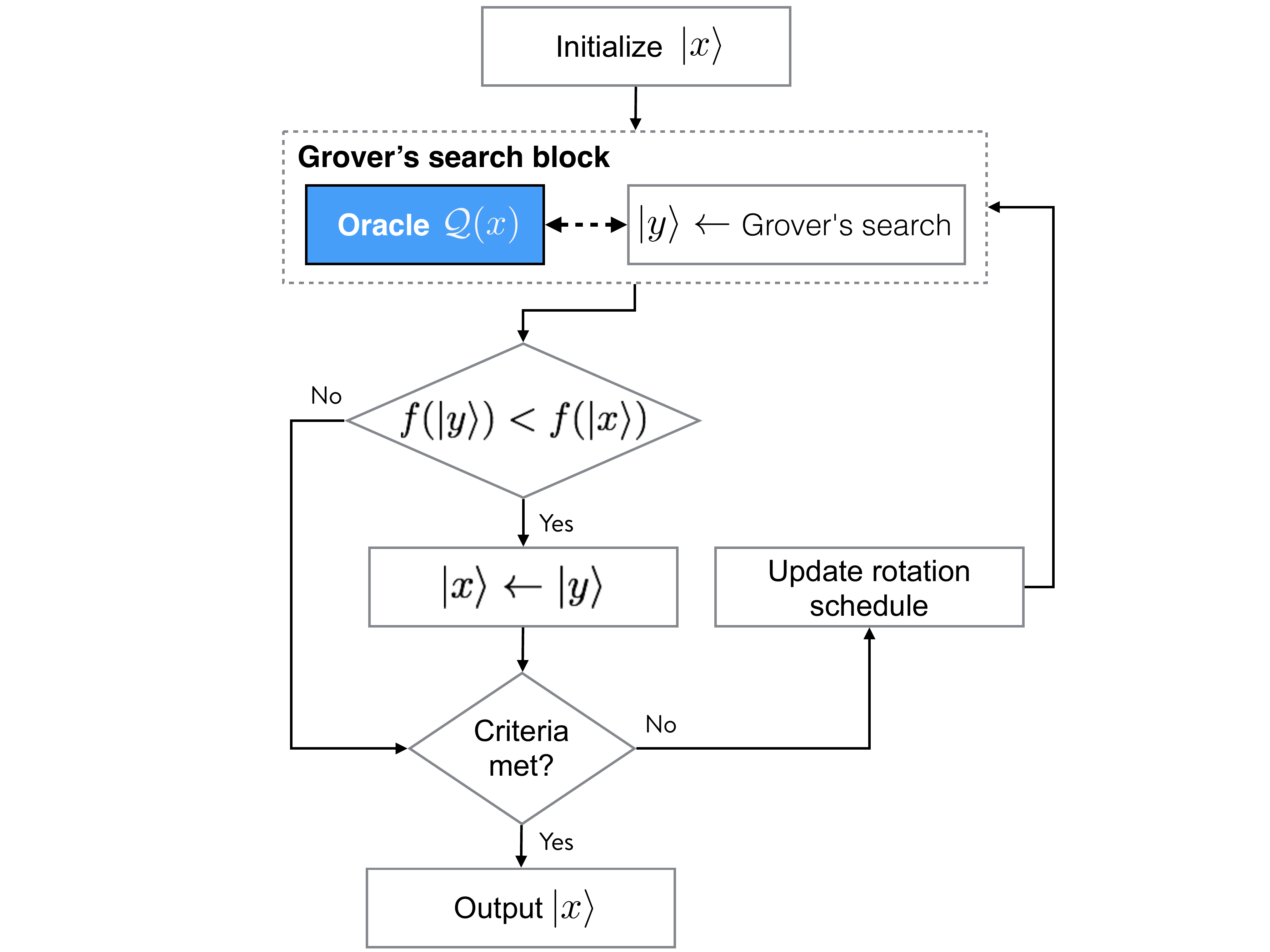}
\caption{A flowchart denoting the general structure of a quantum global optimization algorithm that iteratively employs Grover's search algorithm as a subroutine. This subroutine is referred to as Grover's search block. Note that at each iteration we have to update the structure of the oracle to represent the action of (\ref{eq:grover}).}
\label{fig:flowchart}
\end{figure}

We are now ready to discuss how Grover's search algorithm can be used inside a global optimization algorithm. Fig.~\ref{fig:flowchart} summarizes the general structure of global optimization algorithms which employ Grover's algorithm as part of their structures. The process starts with an initial solution (which is either an incumbent or a random solution). Then, the task of Grover's algorithm is to find any solution which has a lower cost (in terms of the objective function defined in~(\ref{eq:qubo})) than the initial solution. If the resultant search was successful, then we replace the initial solution with a new one and continue the iteration procedure until some criteria are met. In the case that the search process fails, we either abort the algorithm or fine-tune the Grover's search block by using a rotation schedule subroutine. We clarify the role of the schedule generator later in our survey.

Assuming that the implementation of each operator involved in Grover's search algorithm is only polynomial, then the algorithmic complexity of Grover's algorithm to find $\ket{c}$ is $\mathcal{O}(\sqrt{N})$. In \textbf{Algorithm 1}, we have summarized the above geometrical explanation into an algorithmic framework. Here, the task is to find the variable $x_0$ with the fewest queries to the oracle $\mathcal{Q}(x)$.
\\~\\
\begin{minipage}{\linewidth}
\noindent
\textbf{Algorithm 1}~\cite{Gro96,NC00,Stu11}:
\vspace*{2mm}
\hrule
\vspace*{2mm}
\noindent
\textbf{Inputs:} A set of unstructured databases with integer-indexed elements

\noindent
\textbf{Output:} $x_0$

\noindent
\textbf{Resources}:
\begin{itemize}
\item An $n$-qubit register initialized at $\ket{0}^{\otimes{n}}$
\item An oracle in the form of (\ref{eq:oracle})
\end{itemize}

\noindent
\textbf{Runtime:} $\mathcal{O}(\sqrt{2^n})$ operation~\cite{Amb04}, with $\mathcal{O}(1)$ probability of success

\noindent
\textbf{Procedure:}
\begin{enumerate}
\item Initialize the state of the quantum register at $\ket{0}^{\otimes{n}}$
\item Apply Hadamard gate to generate state $\ket{\psi}={\bf H}^{\otimes{n}}\ket{0}^{\otimes{n}}=\frac{1}{2^n}\sum_{x=0}^{2^n-1}\ket{x}$
\item Apply the operators $\bf{DV}$ $R$ times
\item Measure the register
\end{enumerate}
\hrule
\vspace*{2mm}
\end{minipage}

When the search problem is known to have multiple solutions (target states), that is, $c\in\{\mathcal{M}\}$ with $t=|M|$, then the number of iterations needed to find the solution is reduced to $R\approx{\sqrt{N/t}}$~\cite{Liu2010620}. We can explain this using the previous geometrical picture of Grover's algorithm. When there are multiple target states (i.e., there exists degeneracy at the target subspace), we can still represent the initial state, $\ket{\psi}$, in the two-dimensional space of target and non-target states. However, the target state will be a superposition of all of the possible target states $\ket{c}=\frac{1}{\sqrt{t}}\sum_{i\in{t}}\ket{c_i}$. Now, we see that the overlap of the initial state and the $\ket{c_\perp}$ is
\begin{equation}
\label{eq:multi_target}
\theta=\sqrt{\frac{t}{N}}\,.
\end{equation}
From (\ref{eq:multi_target}), we conclude that in order to rotate the initial state toward its $\ket{c}$ component, one should apply $\bf{DV}$ to the initial state $\frac{\pi}{4}\sqrt{N/t}$ times.

We can use Grover's algorithm as a subroutine in a global optimization framework to find the ground state of a QUBO problem. Given the formulation of (\ref{eq:qubo}), the algorithmic procedure reads as follows.
\\~\\
\begin{minipage}{\linewidth}
\noindent
\textbf{Algorithm 2}~\cite{BBW05,BBH+98,Liu2010620}
\vspace*{2mm}
\hrule
\vspace*{2mm}
\noindent
\textbf{Inputs:}
\begin{itemize}
\item An initial trial solution $x^{(0)}$
\item An objective function in the form of $(\ref{eq:qubo})$
\end{itemize}

\noindent
\textbf{Resources:} 
\begin{itemize}
\item A quantum oracle $\mathcal{Q}$ which performs the operation

\begin{equation}
\label{eq:grover}
\mathcal{Q}\ket{x}=(-1)^{\sigma({f(x)-f(x^{(0)}))}}\ket{x},
 \end{equation}
 for all $0\le{x}<2^n$, where $\sigma$ is the sigmoid function.
\item An $n$-qubit register initialized at $\ket{0}^{\otimes{n}}$
\end{itemize}

\noindent
\textbf{Output:} $x_{\text{best}}$

\noindent
\textbf{Runtime:} $\mathcal{O}(\sqrt{N})$ operation~\cite{Amb04}, with $\mathcal{O}(1)$ probability of success

\noindent
\textbf{Procedure:}

\noindent
\begin{itemize}
\item Randomly choose a trial solution, $x^{(0)}$, and initialize $x_\text{best}=x^{(0)}$

\item for $i=1, 2, 3, \dots$ until the convergence criteria are met
\begin{enumerate}
\item Apply \textbf{Algorithm 1} on the register, with the oracle constructed as (\ref{eq:grover})
\item Measure the register and store the output in register $y$
\item Perform a classical evaluation: if ($f(y)<f(x_{\text{best}})$), then $x_{\text{best}}=y$
\end{enumerate}
\end{itemize}
\hrule
\vspace*{2mm}
\end{minipage}

In \textbf{Algorithm 2}, at each iteration, we divide the entire search space into two subspaces. The first subspace includes any state with an energy (in terms of the Ising model Hamiltonian) less than the energy of the incumbent. The second subspace includes the incumbent and any other state which has an energy higher than the energy of the incumbent. Then, the role of Grover's search algorithm in \textbf{Algorithm 2} is to find a solution in the first subspace. At each iteration, we need to update the structure of the oracle if a new incumbent is found. We leave the details of constructing the oracle of a logical function such as $\sigma{(f(x)-f(x_\text{best}))}$  to other works. For example, in~\cite{NC00}, one can see how the quantum circuit of a logical function can be built; however, this does not guarantee the efficiency of the oracle.

In order for Grover's algorithm to show a quadratic speed-up at each iteration, the number of solutions with a lower energy value than the current incumbent must be known. A larger number of rotations does not necessarily guarantee the convergence of the algorithm toward a better solution than the current incumbent. This can be seen from Fig.~\ref{fig:diagram2}. If the size of the target, $t$, is not known in an instance of employing Grover's search algorithm, we do not know the exact overlap between the initial state and the target space. This lack of knowledge prevents us from knowing the exact number of rotations needed to steer the initial state toward the target space. A greater number of rotations than necessary would rotate the initial state away from the $\ket{c}$ axis, resulting in a small probability of finding any target state. Fortunately, we are not completely blind in this case, as there exist algorithms for gaining some insights about the size of the target space.

Given the problem of finding an estimate of the size of the target subspace at each iteration of \textbf{Algorithm~2}, there are a few heuristic and deterministic algorithms proposed in~\cite{BBW05,BBH+98,Liu2010620} for this task. In what follows, we discuss each of these approaches and explain their features.

In~\cite{BBH+98}, a simple heuristic approach is proposed for choosing the number of rotations at each iteration of \textbf{Algorithm~2}. We explain how this heuristic algorithm works at each iteration of the algorithm. First, consider a real number $m^{(k)}$, where the superscript $k$ denotes the value of $m$ at iteration $k$. Then, at the $k$-th iteration, we always choose the number of rotations, $r^{k}$, to be a uniform random number chosen from $[0,m^{(k)}]$. The value of $m^{(k)}$ changes based on the relation
\begin{equation}
\label{eq:heuristics}
    m^{(k)}= 
\begin{cases}
    m^{(k+1)},& f(Y)<f(X_{\text{best}})\\        \text{rand}(m^{(k)}\lambda,\sqrt{N}),         & \text{otherwise}\,;
\end{cases}
\end{equation}
where the $\textbf{rand}$ function generates random numbers and $\lambda\in\{1,4/3\}$ with an optimal value of $\lambda^{*}=6/5$.

A different deterministic approach is proposed by Baritompa et al.~\cite{BBW05}, which involves generating a schedule for the number of rotations in each iteration.  Interestingly, the rotation schedule is independent of the problem instance and works well for different sets of problems given in~\cite{BBW05}. This method assumes a uniform prior over the unknown ratio of $p=t/M$, and updates this distribution independent of success or failure of Grover's search at each iteration. Assuming that $u_i(y)$ is the cumulative distribution function of the improving fraction $p$ at the$k$-th iteration of \textbf{Algorithm~2}, the $r$-schedule generator at each step maximizes an objective function that is the ratio of the benefit, $Eu_i-Ew$, to the cost, $r+1$, with $Eu_i$ and $E_w$ being the expectation under the respective distributions $u_i(y)$ and $u(y)$ of the improving fraction of the domain.

There are other heuristic algorithms which are inspired by~\cite{BBW05}. One such algorithm, which we call the static version of~\cite{BBW05}, is based on having a user-defined prior over the initial distribution of the improving fraction $p$. There is also a dynamic version of~\cite{BBW05} in which the cumulative distribution over the improving fraction is updated differently according to a bias law and the failure or success of Grover's algorithm at each iteration. This dynamic $r$-schedule generator is a preferred approach, as it has to be run on a classical computer at each iteration of \textbf{Algorithm~2}, slowing down the entire optimization procedure. 

We note that although the above $r$-schedule generators reduce the number of rotations each time Grover's search block is employed (i.e., they reduce the number of applications of $\textbf{DV}$), they do not change the computational complexity of Grover's algorithm into an algorithm with polynomial complexity. An estimation of the size of the target subspace, $t$, is required in order to keep the complexity sub-exponential; otherwise, the algorithm turns into a random search over the search space of (\ref{eq:qubo}).

Thus far, what we have discussed was based on the use of Grover's search algorithm. This requires knowledge of the improving fraction $p$ throughout the optimization procedure. There is another class of quantum global optimization algorithms which are inspired by AQC. We discuss one of these algorithms in the following section.

\subsection{A quantum search heuristic}
From our discussion in Section~\ref{sec:AQCQUBO}, we can consider AQC as a process that adiabatically evolves an initial quantum state toward a specific final state that is the ground state of the problem Hamiltonian. One can use this idea and construct a sequence of unitary operations such that the overall action of this sequence on an initial state leads to a final state close to the ground state of an Ising model. As is proposed in~\cite{HP00}, such a sequence of unitary operations can be constructed where the elements of these operations depend on the cost (i.e., the value) of the objective function for a given state. In what follows, we explain the procedure of this algorithm.

The quantum heuristic search algorithm can be generally defined as the sequence of unitary operators
\begin{equation}
\label{eq:Todequ}
{\bf{M}}={\bf{U}}^{(j)}{\bf{P}}^{(j)}\dots {\bf{U}}^{(1)}{\bf{P}}^{(1)},
\end{equation}
where the operator $\bf{M}$ acts on the initial state represented in the form of (\ref{eq:spstate}). In (\ref{eq:Todequ}), at the $k$-th step, the operator ${\bf U}^{k}$ and ${\bf P}^k$ are the mixing and phase operators, respectively, and are defined in what follows. Note that, since the QUBO problems correspond to $k$-local Hamiltonians, we need only a polynomial number of quantum logical gates to construct the operator ${\bf U}^{k}$.

The operator ${\bf U}^{k}$ does not depend on the problem instance. Each element of operator ${\bf U}^{k}$ at a given row ($r$) and column ($c$) is defined as
\begin{equation}
\label{eq:hau1}
{\bf{U}}_{rc}^{k}\equiv{-i\text{tan}(\pi\tau_k/2)}\,.
\end{equation}
The operator ${\bf P}^{k}$ is a diagonal matrix, with each element depending on the cost of the objective function $f(x)$. The $s$-th element of this operator is defined as 
\begin{equation}
\label{eq:hau2}
{\bf{P}}^{(k)}_{ss}=\text{e}^{i\pi\rho_{k}f(x)}\,.
\end{equation}

In both Equation (\ref{eq:hau1}) and Equation (\ref{eq:hau2}), the parameters $\tau_k$ and $\rho_k$ are real numbers that vary linearly at each step throughout the evolution. This linear variation has been a common property used in other classical heuristics as well~\cite{ZSS14}. Applying operator $\bf{M}$ evolves the initial state (\ref{eq:spstate}) toward a minimum energy state of a QUBO problem which is encoded in the Ising spin model Hamiltonian.

It is easy to see that the algorithm described in this section has the same overall structure as amplitude amplification. In particular, we can reduce the current algorithm into an amplitude amplification algorithm~\cite{BHM+02}. However, since in an optimization problem the set of solutions is not known a priori, the amplitude amplification algorithm gives no improvement over the states of the solutions. The fact that the proposed heuristic algorithm depends on the cost of the problem instance precludes a direct asymptotic analysis of the algorithm. 

It is interesting to see that operator $\bf{M}$ has the same form as the the sequence of  unitary operations resulting from the discretization of the adiabatic quantum Hamiltonian in~\cite{DMV01}, as well as the sequence of rotations used in Grover's algorithm described in~\cite{Gro96}. In the following section, we discuss how AQC inspires researchers to devise optimization algorithms that run on gate model quantum computers.

\section{Quantum algorithms to approximate the ground state of the Ising model}
\label{sec:approxi}
In this section, we focus on quantum algorithms that provide an approximation of the ground state. This set of algorithms is based on the introduction of some unknown parameters into the unitary evolution of the system and the subsequent finding of the optimal value of these parameters so as to maximize the overlap between the resultant quantum state under the evolution and the true ground state of the system.

\subsection{Quantum approximate optimization algorithm (QAOA)}
A quantum algorithm to approximate the ground state of a $k$-local Hamiltonian is proposed by Farhi~et~al.~\cite{FGG14}. This algorithm can be used to approximate the ground state of an Ising model. The QAOA has been suggested as an excellent candidate to run on near-term quantum computers, not only because it may be of use for optimization, but also because of its potential as a route to establishing quantum supremacy~\cite{FH16}.

The QAOA is inspired by AQC and works based on the parametrization of the unitary evolution of the quantum system. Given the near-optimal values of the parameters, a sequence of unitary operators acting on the initial state of the system will generate a final state close to the ground state. To explain the procedure of the algorithm, let us consider an Ising model Hamiltonian with at most $k$-local interactions between qubits as follows~\cite{FGG14}:
\begin{equation}
\label{eq:Isingmodel}
\hat{H}=\sum_{k}{H_k}\,.
\end{equation}

The algorithmic steps of the QAOA read as follows:

\begin{enumerate}
\item Generate the initial state as a uniform superposition of all states in the computational basis: $\ket{\psi}_{\text{i}}=H^{\otimes{n}}\ket{0}^{\otimes{n}}$.
\item Construct the unitary operator $U(\hat{H},\gamma)$ which depends on the angle $\gamma$ as follows:
\begin{equation}
U(\hat{H},\gamma)=\text{e}^{-i\gamma\hat{H}}=\prod_{\alpha=1}^m{\text{e}^{-i\gamma{H_\alpha}}}.
\end{equation}
\item Construct the operator $B$ which is the sum of all single-bit $\sigma^x$ operators:
\begin{equation}
B=\sum_{j=1}^n\sigma_j^x.
\end{equation}
\item Define the angle-dependent quantum state for any integer $p\ge{1}$ and $2p$ angles $\gamma_1\dots\gamma_p\equiv\pmb{\gamma}$ and $\beta_1\dots\beta_p\equiv\pmb{\beta}$ as follows:
\begin{equation}
\label{eq:state}
\ket{\pmb{\gamma},\pmb{\beta}}=U(B,\beta_p)U(\hat{H},\gamma_p)\dots{U(B,\beta_1)}U(\hat{H},\gamma_1)\ket{\psi_0}.
\end{equation}
\item Obtain the expectation of $\hat{H}$ in this state (this step could be performed on a quantum computer),
\begin{equation}
\label{eq:ExpH}
F_p(\pmb{\gamma},\pmb{\beta})=\bra{\pmb{\gamma,\beta}}\hat{H}\ket{\pmb{\gamma,\beta}}\,.
\end{equation}
\item Update the parameters ($\pmb\gamma,\pmb\beta$) using a classical (or quantum) optimization algorithm in order to minimize $F_p$.
\item Iterate over steps 5 and 6 in order to find the minimum value of $F_p$ for the near-optimal values ($\pmb\gamma^{*},\pmb\beta^{*}$).
\item Plug ($\pmb\gamma^{*},\pmb\beta^{*}$) into Equation (\ref{eq:state}) and evolve the initial state of the system to the state $\ket{\pmb\gamma^{*},\pmb\beta^*}$.
\item Repeat step 8 with the same angles. A sufficient number of repetitions will produce a state which represents a close enough solution to the ground state of $\hat{H}$.
\end{enumerate}

The above algorithm can be used to approximate the ground state of an Ising model. However, the challenge is in that how to come up with the near-optimal values for the angles is not obvious. The approaches to calculating the optimal values are given in~\cite{FGG14} when one considers the $p$ as a fixed or varying values.
The QAOA inspired other researchers to propose similar ideas for approximating the ground state of the Ising model~\cite{WHT16,YRS+16}.

\subsection{Quantum control and variational methods}
Optimal control theory~\cite{Bry75,Ste12}, combined with variational quantum algorithms~\cite{JJR+16}, is another approach to finding the ground state of an Ising model. In this approach~\cite{YRS+16}, the Hamiltonian of the system is represented by
\begin{equation}
\label{eq:bangH}
\hat{H}=[1-g(t)]H_\text{i}+g(t)H_\text{f}\,,
\end{equation}
where $H_\text{i}$ represents a transverse field generating the quantum fluctuation and $H_\text{f}$ contains the problem Hamiltonian. Here, $0\le{g(t)}\le{1}$ is a time-dependent function. Using the Pontryagin's minimum principle of optimal control theory~\cite{Pon87}, Yang et al. has shown~\cite{YRS+16} that throughout the evolution of the system, the optimal shape of $g(t)$ is bang-bang, that is,~$g(t)$ abruptly changes between the lower and upper bounds of the control parameters. With the knowledge that optimal control policy~\cite{PWZ+16} has a bang-bang shape, one can maximize the following objective function over $g(t)$ to find the optimal path of $g(t)$ for the evolution:
\begin{equation}
\label{eq:bangobj}
F(g(t),T)=\braket{\psi(T)|H_\text{f}|\psi{(T)}}\,.
\end{equation} 
In (\ref{eq:bangobj}), $\ket{\psi(T)}=U(T)\ket{\psi(0)}$, with $U(T)$ as defined in~(\ref{eq:Ue}) and where $T$ is the total evolution time. The objective function's evaluation can be done on a quantum computer, whereas the classical computing device updates the control policy. Given the optimal control policy  $g^*(t)$, one can simulate the evolution of  the system  from $\ket{\psi(0)}$ to $\ket{\psi(T)}$ on a gate model quantum computer. Repeating the evolution process multiple times, and performing measurements in the computational basis of the Ising Hamiltonian will lead to a state which has a large overlap with the true ground state of the system.

There is a direct connection between the above control procedure and the QAOA. Since it has been mathematically shown in~\cite{YRS+16} that the optimal schedule takes a bang-bang shape, one can parametrize the above procedure in a way that is similar to the QAOA and use the time interval of discretized pulses as variational parameters to be optimized using a classical optimization algorithm, while increasing $p$ to achieve convergence.

\subsection{Training a quantum optimizer}
Wecker et al.~\cite{WHT16} have transformed the problem of approximating the ground state of an Ising model into a machine learning task. In particular, they turn the problem into a supervised machine learning problem, where the training set is a subset of hard instances studied in~\cite{CFL+14}. Knowing the ground state of these hard instances, they generate specific schedules which give the highest overlap between the ground state and the final state after the evolution. Technically, the objective function is rather different than Equation (\ref{eq:ExpH}) and is represented as follows:
\begin{equation}
\label{eq:DaceWecker}
F_p(\pmb\theta_Z,\pmb\theta_X)=|\braket{\pmb\theta_X,\pmb\theta_Z|\psi_g}| \,,
\end{equation}
where $\ket{\psi_g}$ is the true ground state and $(\pmb\theta_X,\pmb\theta_Z)$ are the learning parameters which determine the evolution of the system from the initial state to $\ket{\pmb\theta_X,\pmb\theta_Z}$. The training procedure was performed on instances with $n=20$ variables, but tested on sizes up to $N=28$, where it shows a larger overlap for these instances than the optimized annealing times proposed in~\cite{CFL+14}. We should note that the approach proposed by Wecker et al. works well in the case of Max-SAT (maximum satisfiability) problems. Since it is shown in~\cite{bian2010ising} how to transform a QUBO problem into a weighted Max-SAT problem, it remains to be seen how the method in~\cite{WHT16} can be generalized to solve weighted Max-SAT problems (i.e., general QUBO problems).

\section{Quantum Metropolis sampling}
\label{sec:qms}
Sampling from low energy levels of the quantum system enables algorithmic approaches to find the ground state of a physical system. For instance, good samples from the low energy levels of the system can help to quickly identify the backbone of combinatorial optimization problems by reducing the original optimization problem into smaller subproblems~\cite{KR16,KGK17}. There are other algorithms that benefit from low energy state distributions that have applications in machine learning~\cite{CLG+16,AAR+16}. One can also use sampling to estimate the ground state of an Ising model Hamiltonian. The idea is to initiate the system at some random initial state and then use a quantum sampling algorithm to evolve the system toward a state with lower energy. One example of a sampling algorithm is Metropolis sampling.

The classical Metropolis algorithm is used ubiquitously in the field of computational physics. By evolving the physical system toward its ground or Gibbs state, one can simulate the equilibrium or static properties of a classical system. The corresponding quantum algorithm that performs a Metropolis sampling on the quantum system is proposed in~\cite{TOV+11}. Impervious to the negative-sign problem in classical simulations of fermionic systems, the quantum Metropolis algorithm can simulate the equilibrium and static properties of a wide range of physical systems, including  classical Ising models.

The basic concept of the Metropolis algorithm is to construct a fast-mixing Markov chain which obeys the detailed balance that samples from the states with the highest probabilities. In the case of the classical Ising model, one can start with a random initial state and follow a series of update rules to prepare the system in a state with a lower energy. For an instance of an update rule, we can consider a local random bit-flip operation on the state. If the resultant configuration has a lower energy than the original state, then we accept the move; otherwise, in order to avoid being trapped in a local minimum, we accept the new state with a probability proportional to $\text{exp}(\beta(E_{\text{old}}-E_{\text{new}}))$, where $\beta$ is the inverse temperature and $E$ is the energy of the state.

The quantum version of the Metropolis algorithm acts much in the same way as the classical algorithm; however, the accept/reject step of the quantum version becomes non-trivial in the following sense. At each iteration of the algorithm, we need to perform a measurement on the system to figure out whether we succeeded in achieving a better state (i.e., a state with a lower cost than the one which is already available). In the case of success, we can set the entire system to the new state. In the case of failure, we only accept the new state according to the update rules, and might need to set the system in its previous state. However, we no longer have access to the previous state of the system after performing the initial measurement. Classically, we can always have a copy of the state of the system at any given time; however, quantum mechanically, this is prohibited because of the no-cloning theorem. The quantum Metropolis algorithm resolves this issue through a systematic approach which we now explain.

As usual, readers are invited to consult with the original paper~\cite{TOV+11} for more details. We note that the main component of the quantum Metropolis algorithm is the quantum phase estimation algorithm which works as follows. Let us assume that the action of a quantum phase estimation circuit can be represented by the operator $\mathcal{L}$ which acts on $(n+r)$ qubits, where $n$ and $r$ are the number of qubits of the quantum system and and an extra quantum register, respectively. Given a quantum system in one of its eigenstates $\ket{\phi_i}$ with the corresponding energy $E_i$, the action of $\mathcal{L}$ on the state of the quantum system and register reads as follows:
\begin{equation}
\label{eq:phase}
\ket{\phi_i}\ket{E_i}=\mathcal{L}(\ket{\phi_i}\ket{0})\,,
\end{equation}
where $\ket{E_i}$ is encoded inside the extra quantum register up to $r$-digit precision. Here, through our review of the quantum Metropolis algorithm, we assume a perfect (i.e., error-free) quantum phase estimation procedure. However, it has been shown in~\cite{YRS+16} that the quantum Metropolis algorithm works even when errors are present in the phase estimation algorithm.

The approach explained above can be used to solve for the ground state of an Ising model Hamiltonian. The algorithm starts with a random initial state as an initial solution of the Ising model. We then use the quantum Metropolis algorithm to evolve the initial state toward some states with low energy levels. Sampling from these states with low energies will give us an approximation of the ground state of the Ising model.

The whole quantum Metropolis algorithm acts on the four registers, where we call this action on four registers as a mapping $\mathcal{E}$. The first register, with $n$ qubits, encodes the state of the quantum system; the second register, with $r$ qubits, works as a set of ancilla qubits for the quantum phase estimation algorithm; the third register, with $r$ qubits, holds a copy of the corresponding energy of the initial state of the system at each step of the quantum Metropolis algorithm; and the fourth register, with one qubit, is used for the process of rejection or acceptance of a new state with lower energy than the initial state of the quantum system. The initial state of the first register is an arbitrary state~$\ket{\phi}$. The initial states of all of the other registers are set to $\ket{0}$. Therefore, we represent the initial state of the system as $\ket{\phi}\ket{0}\ket{0}\ket{0}$, showing the states of each of the registers 1--4 in order from left to right. We maintain this ordering in representing the state of the whole registers throughout this section.

The first step in the Metropolis quantum algorithm is to prepare the quantum system in one of its eigenstates that has  a known energy value. This step can be accomplished by preparing the quantum system at some arbitrary state~$\ket{\phi}$, and using the quantum phase estimation algorithm \cite{Kit95} to project the state of the first register into an eigenstate of the system Hamiltonian, while the second register holds the corresponding energy $E_i$. Therefore, the first transformation takes the following form:
\begin{equation}
\label{eq:step1}
\ket{\phi}\ket{0}\ket{0}\ket{0}\longrightarrow\ket{\phi_i}\ket{E_i}{0}\ket{0}\,.
\end{equation}

After generating the new state, we employ a series of $C-NOT$ gates to copy the state of the second register into the third register, and apply, $\mathcal{L}^\dagger$ on the first and second registers. It is easy to show that this sequence of actions leads to the following transformation:
\begin{equation}
\label{eq:step2}
\ket{\phi_i}\ket{E_i}\ket{0}\ket{0}\longrightarrow\ket{\phi_i}\ket{0}\ket{E_i}\ket{0}.
\end{equation}

Similar to the classical Markov chain, at this point we need a transition matrix $\mathcal{C}$ which decides the probability of transition from one state to another. To maintain the ergodicity of the mapping $\mathcal{E}$, one can choose the operator $\mathcal{C}$ to be a universal gate set~\cite{BBC+95}. The quantum detailed balance condition also requires that the probability of applying a specific $\mathcal{C}$ be the same as the probability of applying its conjugate $\mathcal{C}^\dagger$. For instance, a simple choice for $\mathcal{C}$ can be a set of local bit flips at random locations or other simple set of transformations.

The action of the $\mathcal{C}$ operator on the quantum system register $\ket{\phi_i}$ is to take this state into a superposition of all possible eigenstates of the system Hamiltonian, that is, $\mathcal{C}\ket{\phi_i}=\sum_l{c^i_l}\ket{\phi_k}$. Having the action of $\mathcal{C}$ on the first register followed by the action of $\mathcal{L}$ on the first and second registers generates the following mapping:
\begin{equation}
\label{eq:step3}
\ket{\phi_i}\ket{0}\ket{E_i}\ket{0}\longrightarrow\sum_l{c^i_l}\ket{\phi_i}\ket{E_l}\ket{E_i}\ket{0}.
\end{equation}

Arriving at Equation~(\ref{eq:step3}), one might perform a measurement on the second register to get an energy $\ket{E_l}$ and its corresponding eigenstate $\ket{\phi_l}$. If this measurement is successful, then we have a new state with lower energy than the initial energy $\ket{E_i}$, and we can start the entire algorithm over with this new eigenstate. However, in the case of rejection (i.e., $E_l<E_i$) the whole state is destroyed and there is no way to return to the initial state $\ket{\phi_i}$.
This is why we use the fourth register as follows to overcome this problem.

We define the operator $\mathcal{W}$,which is conditioned on the state of the second and third registers (i.e., it depends on the two energies $\ket{E_i}$ and $\ket{E_l}$). This operator performs the following transformation on (\ref{eq:step3}):
\begin{equation}
\label{eq:step4}
\sum_l{c^i_l}\ket{\phi_i}\ket{E_l}{E_i}\ket{0}\longrightarrow\underbrace{\sum_l{c^i_l}\sqrt{h^i_l}\ket{\phi_k}\ket{E_i}\ket{E_l}\ket{1}}_\textrm{$\ket{\psi}_i^{\bot}$}+\underbrace{\sum_l{c^i_l}\sqrt{1-h^i_l}\ket{E_i}\ket{E_l}\ket{0}}_\textrm{$\ket{\psi}_i^{\parallel}$}\,\,\,,
\end{equation}
where $h_l^i=\text{min}(1,\text{exp}(-\beta(E_i-E_l)))$, with $\beta$ representing the inverse temperature. Note that in the above formula, the amplitude $c^i_lh^i_l$ corresponds exactly to the transition probability of the classical Metropolis rules. At this step, one can perform a measurement on the fourth register. An output of $\ket{1}$ is indicative of a new state with a lower energy than the initial state. We then perform a projective measurement on the third register to obtain a new state with the energy $\ket{E_l}$ with its corresponding eigenstate encoded in the first register. However, an output of $\ket{0}$ after measuring the first register indicates a rejection, in which case we must return to the initial state.

We can use the same approach that is employed in~\cite{MW05} to return to the initial state or a state with the same energy as the initial state. We first explain how this measurement works and then we represent the procedure inside a mathematical framework. From (\ref{eq:step4}), we can see that prior to performing any measurement on the fourth register, the entire system is in a linear combination of two orthonormal states \{$\ket{\psi}_i^{\bot}, \ket{\psi}_i^{\parallel}$\}. After performing the first binary measurement on the first register, the state of the system will collapse into one of these subspaces. Independent of the output of the first measurement, we perform a second measurement acting on the two-dimensional space of \{$\ket{\phi}_i, \ket{\phi}_i^{\bot}$\}. This second measurement collapses the current state of the system into the one of these two subspaces. The second von Neumann measurement can easily be implemented using a phase estimation algorithm that compares the energy of the current state with the energy of the initial state which is encoded in the third register. If the result of the second measurement is $E_i=E_k$, then we have successfully prepared the system in a quantum state with the same energy as the initial state $\ket{\phi}_i$. Otherwise, we repeat this measurements sequence until we retrieve the initial state.

The sequence of binary measurements explained in the previous paragraph take the state of the system into two different two-dimensional spaces represented by different bases. Beginning with the first measurement performed on the first qubit, we can define the complete set of projectors $\mathcal{Q}_\bot$ and $Q_\parallel$ as follows:
\begin{align}
\label{eq:bin_proj1st}
\mathcal{Q}_0=\ket{0}\bra{0}, \\ \nonumber
\mathcal{Q}_1=\ket{1}\bra{1}.
\end{align}
The above projective measurement will collapse the system into an accepted state {$\ket{\psi}_i^{\parallel}$ if the resultant measurement is $\mathcal{Q}_1$; otherwise, it collapses the system into a rejected state {$\ket{\psi}_i^{\bot}$. Since we have a complete set of projector operators $\mathcal{Q}=\mathcal{Q}_0+\mathcal{Q}_1=\mathds{1}$, we can expand each of the states $\ket{\phi}_i$ or $\ket{\phi}_i^{\bot}$} in the \{$\ket{\psi}_i^{\bot}, \ket{\psi}_i^{\parallel}$\} basis as follows:
\begin{equation}
\label{eq:firstmeasure}
\begin{aligned}
&\ket{\phi}_i  =\mathcal{Q}\ket{\phi}_i=\sqrt{p}\ket{\psi}_i^\parallel +\sqrt{1-p} \ket{\psi}_i^\bot \\
&\ket{\phi}_i^\bot =\mathcal{Q}\ket{\phi}_i^{\bot}=\sqrt{1-p}\ket{\psi}_i^\parallel -\sqrt{p} \ket{\psi}_i^\bot.
\end{aligned}
\end{equation}
The second set of projective measurements collapse the state of the system into one of the subspaces $\ket{\phi}_i$ or $\ket{\phi}_i^\bot$. As mentioned earlier, we can implement this projective measurement on the original subspace by using a phase estimation technique on the second and third registers. The mathematical representation of the these projection operators can be stated simply as
\begin{align}
\label{eq:bin_proj2st}
\mathcal{P}_0=\ket{\phi}_i\bra{\phi}_i, \\ \nonumber
\mathcal{P}_1=\ket{\phi}_i^{\bot}{\bra{\phi}_i^{\bot}}.
\end{align}
Following the measurements stated in (\ref{eq:bin_proj2st}), if the outcome of the measurement is $\mathcal{P}_i$, then we have successfully returned to $\ket{\phi}_i$; otherwise, the system collapses into the state $\ket{\phi}_i{^\bot}$.

It is easy to derive an inverse transformation from (\ref{eq:firstmeasure}) and express the states \{$\ket{\psi}_i^{\bot}, \ket{\psi}_i^{\parallel}$\} in terms of the orthonormal basis \{$\ket{\phi}_i$, $\ket{\phi}_i^{\bot}$\} as follows:
\begin{equation}
\label{eq:secondmeasure}
\begin{aligned}
&\ket{\psi}_i^\parallel = \sqrt{p}\ket{\phi}_i +\sqrt{1-p} \ket{\phi}_i^\bot \\     
&\ket{\psi}_i^\bot = \sqrt{1-p}\ket{\phi}_i -\sqrt{p} \ket{\phi}_i^\bot.
\end{aligned}
\end{equation}
Assuming that the system is in a rejection state, Fig.~\ref{fig:diagram1} demonstrates the recurrence of rejection  through the sequence of measurements $Q_\alpha$$P_\alpha$. It is proved that the probability of reaching a state with the same energy as the initial state $\ket{\phi}_i$ increases exponentially with the number of measurements.

\begin{figure}
\centering
\includegraphics[trim ={0 10cm 0 9cm},clip, width=\textwidth]{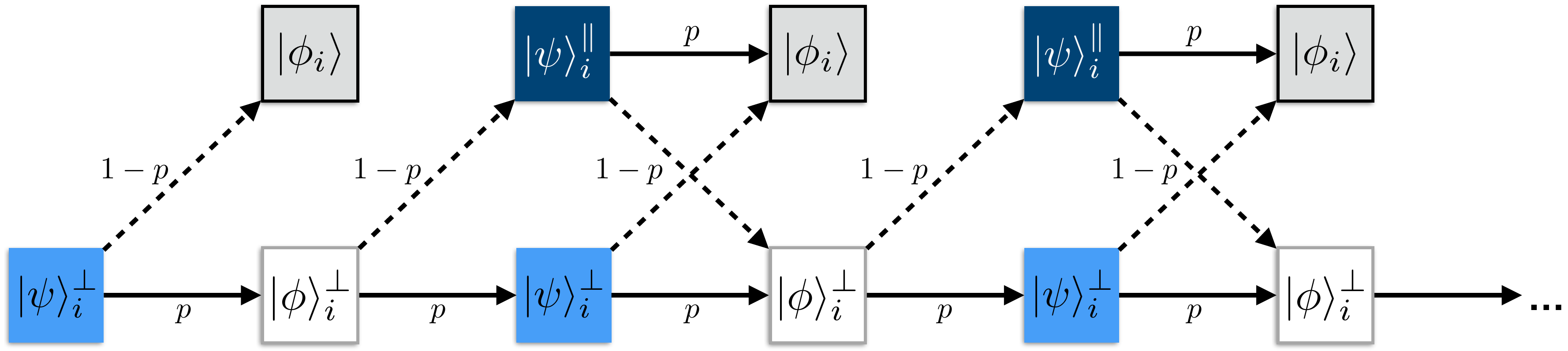}
\caption{A procedure representing recurrence of rejection, followed by the sequence of measurements $Q_\alpha$$P_\alpha$. The recurrence procedure terminates when it reaches  the state $\ket{\phi}_i$.}
\label{fig:diagram1}
\end{figure}

\section{Quantum algorithms for solving specific types of QUBO problems}
\label{sec:specific}
Thus far, we have only discussed the quantum algorithms that can solve a general QUBO problem on a gate model quantum computer, without any guarantee that the resultant quantum algorithms are faster than the best classical algorithms. There exist quantum algorithms for specific types of QUBO problems, including (but not limited to) satisfiability problems~\cite{Amb04}, the triangle finding problem~\cite{MSS07,Gal14}, finding a clique with a length of $k$~\cite{CE03}, and minimum spanning trees and graph connectivity~\cite{DHH+04}. Most of these algorithms use Grover's search or quantum amplitude amplification algorithm~\cite{BHM+02} to devise a quantum algorithm that is faster than the best classical alternative}. In the following sections, we discuss three quantum algorithms for solving constraint satisfaction problems~\cite{CGW00}, Boolean satisfiability problems~\cite{Amb04}, maximum independent set problems~\cite{Dor05}, and clustering problems~\cite{LMR13}.

\subsection{Constraint satisfaction problems}
Constraint satisfaction problems (CSP) are defined as a set of variables, each taking a finite set of domain values, which together are involved in a set of constraints. A set of values must be assigned to the variables such that no constraints are violated. Cerf et al. propose a nested quantum algorithm for solving CSPs, which exploits the quadratic speed-up of Grover's search algorithm. As is explained in~\cite{CGW00}, the algorithm can be summarized as follows:

\begin{enumerate}
\item Construct a binary tree representing the CSP.
\item Construct a superposition of all possible solutions at level $i$ of the tree.
\item Use the quantum search algorithm to amplify the magnitude of could-be solutions using  \textbf{Algorithm~1}.
\item Perform a subsequent quantum search in the subspace of the descendants of all of the could-be partial solutions simultaneously. This is achieved by using \textbf{Algorithm~1} as an input, the superposition of could-be solutions resulting from the previous step. The overall yield after this step is a superposition of all states where the solutions have been partially amplified with respect to non-solutions.
\item The final procedure consists of nesting stages 3 and 4, using them as a search operator inside a higher-level quantum search algorithm until the solutions get maximally amplified, at which point a measurement is performed.
\end{enumerate}

For a pseudo-Boolean CSP, the number of iterations needed to find a solution of an average instance of the problem scales as $\mathcal{O}(2^{\alpha{n}})$, with $\alpha<1$, depending on the nesting depth and problem type.

\subsection{Satisfiability problems}
Although a Boolean satisfiability problem, known as the SAT problem, can be thought of as a certain form of a conjunctive normal form (CNF) problem, there exist specific quantum algorithms for solving SAT problems. In a SAT problem, one is given a Boolean function $F$ in the CNF
\begin{equation}
\label{eq:Boolean}
F=C_1\wedge{C_2}\dots\wedge{C_m}\,,
\end{equation}
where $m$ is the number of clauses, each expressed as the disjunction of literals, $l_i$, as
\begin{equation}
C_j=l_1\vee{l_2}\dots\vee{l_k}\,.
\end{equation}
The task with satisfiability problems is to find an assignment that makes $F$ true. A general SAT problem is called a $k$-SAT problem if the number of literals in each clause is restricted to $k$. A SAT problem is a special case of a weighted Max-SAT (WMSAT) problem. In a WMSAT problem, one is given a Boolean function such as (\ref{eq:Boolean}) with each clause expressed as $C_j=(l_1\vee{l_2}\dots\vee{l_k},\omega_j)$, where each weight $\omega_j$ is a positive number. The objective is to maximize the sum of the weights of satisfied clauses by any assignment. 

One of the fastest classical SAT solvers, proposed by Sch\"{o}ning, solves a $k$-SAT problem in \mbox{$\mathcal{O}((2-2/k)^n)$}~\cite{Sch99}. It would be interesting to see whether a quantum algorithm exists that performs better than the proposed algorithm in~\cite{Sch99}. In fact, Ambainis~\cite{Amb04} proposed a quantum algorithm that solves a $k$-SAT problem in $\mathcal{O}((2-2/k)^{\frac{n}{2}})$. The algorithm employs the amplitude amplification technique~\cite{BHM+02} that is a generalization of Grover's algorithm. Amplitude amplification techniques can even be used to gain a quadratic speed-up over a classical SAT solver that has the best upper bound~\cite{PPZ97,PPS+98}. The preceding argument also holds for the best classical SAT solver (with no constraint on the clauses) such that the resultant quantum algorithm provides a quadratic speed-up over the best classical algorithm~\cite{Ber05}. In summary, it has been shown in~\cite{DKW05} that by using variants of Grover's search algorithm one cannot attain better than quadratic speed-up over the best classical algorithms for solving SAT problems.
 
 \subsection{Independent set problems}
 An independent set problem is the problem of finding the set of vertices of a graph under the constraint that no two of the vertices can be adjacent. Finding the maximum number of these vertices under this constraint is a maximum independent set (MIS) problem. An MIS problem belongs to the class of NP-complete problems~\cite{BK94,Joh79}. Thus far, the best classical algorithm has a time complexity of $\mathcal{O}(1.1844^n)$~\cite{Rob01}, where $n$ is the number of vertices. D\"{o}rn proposes a quantum algorithm that employs the quantum amplitude amplification algorithm for finding the MIS with a time complexity of $\mathcal{O}(1.1488^n)$~\cite{Dor05}.

\subsection{Data clustering in supervised and unsupervised machine learning}
Lloyd et al. \cite{LMR13} show that a quantum algorithm can be improved upon using Grover's search algorithm to solve the supervised learning $k$-means clustering problem with an exponential speed-up. They further discuss that the unsupervised version of this problem can be reformulated as a QUBO problem that is amenable to the adiabatic algorithm. According to our discussion above, we can extend this approach to work on a gate model quantum hardware device by using any of the ideas for solving QUBO problems on a gate model quantum computer that are mentioned in this survey. There are many other works in the literature that attempt to address various machine learning problems using quantum algorithms (see, for instance,~\cite{Wittek14}). \\

In this section, we have discussed a few application-specific quantum algorithms for solving certain families of combinatorial optimization problems. There are many other examples in the literature, and creating a comprehensive list of the various ideas is outside of the scope of this research. Our main purpose was to show that, based on the same few principles described earlier (such as the use of Grover's search algorithm or the unified QUBO framework), new custom-tailored algorithms can be developed to address specific problems tied to certain applications.

\section{Discussion and conclusion}
\label{sec:conclude}
Quantum computing is a new paradigm of computation that has the potential to  surpass the computing power of classical computers for certain computational tasks. Combinatorial optimization is the large family of discrete-valued optimization problems that includes many such computational tasks that are hard for classical computers. Emerging circuit-model quantum hardware based on scalable technologies has invigorated investigations into applying these hardware devices to solve combinatorial optimization problems.\\

In this survey, we have reviewed the existing gate model quantum algorithms for solving combinatorial optimization problems. In particular, we have investigated the framework of formulating combinatorial optimization problems in the  quadratic unconstrained binary optimization (QUBO) formalism. Whether it is possible to develop a quantum-enabled general QUBO solver that runs on a gate model quantum computer is the question we addressed in this work.

Our survey has discussed some approaches to solving a general QUBO problem on a gate model quantum computer. In the first approach, we introduced the simulation of adiabatic quantum computing (AQC), and quantum annealing in particular, on a gate model quantum computer.  AQC is naturally closer to discrete optimization problems. However, a simulation of AQC on a gate model computer might have the advantage of being able to benefit from more-practical error correction schemes.

Using Grover's search algorithm as a subroutine in global optimization algorithms is another approach to solving QUBO problems. Although this approach is appealing, the computational complexity of this approach is $\mathcal{O}(\sqrt{2^n})$, which is slower in practice than many efficient classical algorithms. There have been numerous algorithms developed that are based on the idea of using Grover's search algorithm as a subroutine. One way to speed up this na\"ive application of Grover's algorithm is to combine it with other local optimization algorithms to overcome the curse of dimensionality~\cite{Bul05,Bul07}. There is much room for the improvement and development of more-sophisticated hybrid quantum--classical optimization algorithms based on Grover's search algorithm.

State-of-the-art chip design and fabrication technologies allow us to have quantum processors with a few hundred qubits. Larger chips with more-scalable architectures are expected to become available as an advancement in the longer term. Therefore, the development of quantum algorithms that require small quantum processing units has become an active area of research in quantum computing. The quantum algorithms that can approximate the ground state of the classical Ising model like quantum approximate optimization algorithm and the quantum Metropolis algorithm, can be considered good examples of these algorithms.

This survey has provided example approaches to illustrate that the solving of combinatorial optimization problems using a unified programming framework (such as QUBO) is possible on a gate model quantum computer. Our hope is that a consolidation of ideas such as this can help researchers come up with more-advanced algorithms that can harness the potential of gate model quantum computers to solve hard combinatorial optimization problems.

\section*{Acknowledgment}
The authors would like to thank Marko Bucyk for editing
the manuscript, Robert Raussendorf, and Anna  Levit for useful discussion. This work was supported by 1QBit. 


\begin{thebibliography}{96}
\providecommand{\natexlab}[1]{#1}
\providecommand{\url}[1]{{#1}}
\providecommand{\urlprefix}{URL }
\expandafter\ifx\csname urlstyle\endcsname\relax
  \providecommand{\doi}[1]{DOI~\discretionary{}{}{}#1}\else
  \providecommand{\doi}{DOI~\discretionary{}{}{}\begingroup
  \urlstyle{rm}\Url}\fi
\providecommand{\eprint}[2][]{\url{#2}}

\bibitem[{Ladd et~al(2010)Ladd, Jelezko, Laflamme, Nakamura, Monroe, and
  O'Brien}]{LJL+10}
Ladd TD, Jelezko F, Laflamme R, Nakamura Y, Monroe C, O'Brien JL (2010) Quantum
  computers. Nature 464(7285):45--53

\bibitem[{Lloyd(1996)}]{Llo96}
Lloyd S (1996) Universal quantum simulators. Science 273(5278):1073--1078

\bibitem[{Lanyon et~al(2010)Lanyon, Whitfield, Gillett, Goggin, Almeida,
  Kassal, Biamonte, Mohseni, J., Barbieri, Aspuru-Guzik, and White}]{LWG+17}
Lanyon BP, Whitfield JD, Gillett GG, Goggin ME, Almeida MP, Kassal I, Biamonte
  JD, Mohseni M, J P, Barbieri M, Aspuru-Guzik A, White AG (2010) Towards
  quantum chemistry on a quantum computer. Nat Chem 2(2):106--111

\bibitem[{Shor(1999)}]{Sho99}
Shor PW (1999) Polynomial-time algorithms for prime factorization and discrete
  logarithms on a quantum computer. SIAM Review 41(2):303--332

\bibitem[{Grover(1996)}]{Gro96}
Grover LK (1996) A fast quantum mechanical algorithm for database search. In:
  Proceedings of the Twenty-eighth Annual ACM Symposium on Theory of Computing,
  ACM, New York, NY, USA, STOC '96, pp 212--219

\bibitem[{Nielsen and Chuang(2000)}]{NC00}
Nielsen M, Chuang I (2000) Quantum Computation and Quantum Information.
  Cambridge Series on Information and the Natural Sciences, Cambridge
  University Press

\bibitem[{Farhi et~al(2000)Farhi, Goldstone, Gutmann, and Sipser}]{FGG+00}
Farhi E, Goldstone J, Gutmann S, Sipser M (2000) Quantum computation by
  adiabatic evolution. arXiv preprint quant-ph/0001106

\bibitem[{Shor(1996)}]{Sho96}
Shor PW (1996) Fault-tolerant quantum computation. In: Foundations of Computer
  Science, 1996. Proceedings., 37th Annual Symposium on, pp 56--65

\bibitem[{Benhelm et~al(2008)Benhelm, Kirchmair, Roos, and Blatt}]{JGC+08}
Benhelm J, Kirchmair G, Roos CF, Blatt R (2008) Towards fault-tolerant quantum
  computing with trapped ions. Nat Phys 4(6):463--466

\bibitem[{Clarke and Wilhelm(2008)}]{CW08}
Clarke J, Wilhelm FK (2008) Superconducting quantum bits. Nature
  453(7198):1031--1042

\bibitem[{Li et~al(2003)Li, Wu, Steel, Gammon, Stievater, Katzer, Park,
  Piermarocchi, and Sham}]{Li809}
Li X, Wu Y, Steel D, Gammon D, Stievater TH, Katzer DS, Park D, Piermarocchi C,
  Sham LJ (2003) An all-optical quantum gate in a semiconductor quantum dot.
  Science 301(5634):809--811

\bibitem[{Johnson et~al(2011)Johnson, Amin, Gildert, Lanting, Hamze, Dickson,
  Harris, Berkley, Johansson, Bunyk, Chapple, Enderud, Hilton, Karimi,
  Ladizinsky, Ladizinsky, Oh, Perminov, Rich, Thom, Tolkacheva, Truncik,
  Uchaikin, Wang, Wilson, and Rose}]{JAG+11}
Johnson MW, Amin MHS, Gildert S, Lanting T, Hamze F, Dickson N, Harris R,
  Berkley AJ, Johansson J, Bunyk P, Chapple EM, Enderud C, Hilton JP, Karimi K,
  Ladizinsky E, Ladizinsky N, Oh T, Perminov I, Rich C, Thom MC, Tolkacheva E,
  Truncik CJS, Uchaikin S, Wang J, Wilson B, Rose G (2011) Quantum annealing
  with manufactured spins. Nature 473(7346):194--198

\bibitem[{Childs et~al(2001)Childs, Farhi, and Preskill}]{CFP01}
Childs AM, Farhi E, Preskill J (2001) Robustness of adiabatic quantum
  computation. Phys Rev A 65:012,322

\bibitem[{Amin and Steininger(2006)}]{AM06}
Amin M, Steininger M (2006) Adiabatic quantum computation with superconducting
  qubits. US Patent 7,135,701

\bibitem[{Hernandez et~al(2016)Hernandez, Zaribafiyan, Aramon, and
  Naghibi}]{HZA+16}
Hernandez M, Zaribafiyan A, Aramon M, Naghibi M (2016) A novel graph-based
  approach for determining molecular similarity. arXiv preprint arXiv:160106693

\bibitem[{Zaribafiyan et~al(2017)Zaribafiyan, Marchand, and
  Changiz~Rezaei}]{ZMR+16}
Zaribafiyan A, Marchand DJJ, Changiz~Rezaei SS (2017) Systematic and
  deterministic graph minor embedding for cartesian products of graphs. Quantum
  Information Processing 16(5):136

\bibitem[{Venturelli et~al(2015)Venturelli, Marchand, and Rojo}]{VMR15}
Venturelli D, Marchand DJ, Rojo G (2015) Quantum annealing implementation of
  job-shop scheduling. arXiv preprint arXiv:150608479

\bibitem[{McGeoch and Wang(2013)}]{MW13}
McGeoch CC, Wang C (2013) Experimental evaluation of an adiabiatic quantum
  system for combinatorial optimization. In: Proceedings of the ACM
  International Conference on Computing Frontiers, ACM, New York, NY, USA, CF
  '13, pp 23:1--23:11

\bibitem[{Rieffel et~al(2015)Rieffel, Venturelli, O'Gorman, Do, Prystay, and
  Smelyanskiy}]{RVO+15}
Rieffel EG, Venturelli D, O'Gorman B, Do MB, Prystay EM, Smelyanskiy VN (2015)
  A case study in programming a quantum annealer for hard operational planning
  problems. Quantum Information Processing 14(1):1--36

\bibitem[{Heim et~al(2015)Heim, R{\o}nnow, Isakov, and Troyer}]{HRI15}
Heim B, R{\o}nnow TF, Isakov SV, Troyer M (2015) Quantum versus classical
  annealing of ising spin glasses. Science 348(6231):215--217

\bibitem[{Aharonov et~al(2008)Aharonov, van Dam, Kempe, Landau, Lloyd, and
  Regev}]{ADK+08}
Aharonov D, van Dam W, Kempe J, Landau Z, Lloyd S, Regev O (2008) Adiabatic
  quantum computation is equivalent to standard quantum computation. SIAM
  Review 50(4):755--787

\bibitem[{Kassal et~al(2008)Kassal, Jordan, Love, Mohseni, and
  Aspuru-Guzik}]{KJL+08}
Kassal I, Jordan SP, Love PJ, Mohseni M, Aspuru-Guzik A (2008) Polynomial-time
  quantum algorithm for the simulation of chemical dynamics. Proceedings of the
  National Academy of Sciences 105(48):18,681--18,686

\bibitem[{Kitaev et~al(2002)Kitaev, Shen, and Vyalyi}]{KSV02}
Kitaev AY, Shen A, Vyalyi MN (2002) Classical and quantum computation, vol~47.
  American Mathematical Society Providence

\bibitem[{Lenstra and Pomerance(1992)}]{Len92}
Lenstra HW, Pomerance C (1992) A rigorous time bound for factoring integers.
  Journal of the American Mathematical Society 5(3):483--516

\bibitem[{Pomerance(1982)}]{Pom82}
Pomerance C (1982) Analysis and comparison of some integer factoring
  algorithms. Mathematisch Centrum Computational Methods in Number Theory, Pt 1
  p 89-139(SEE N 84-17990 08-67)

\bibitem[{Hyafil and Rivest(1973)}]{LR73}
Hyafil L, Rivest RL (1973) Graph partitioning and constructing optimal decision
  trees are polynomial complete problems. IRIA. Laboratoire de Recherche en
  Informatique et Automatique

\bibitem[{Garey et~al(1976{\natexlab{a}})Garey, Johnson, and Sethi}]{GJS76}
Garey MR, Johnson DS, Sethi R (1976{\natexlab{a}}) The complexity of flowshop
  and jobshop scheduling. Math Oper Res 1(2):117--129

\bibitem[{Garey et~al(1976{\natexlab{b}})Garey, Johnson, and
  Stockmeyer}]{GJS76_2}
Garey M, Johnson D, Stockmeyer L (1976{\natexlab{b}}) Some simplified
  np-complete graph problems. Theoretical Computer Science 1(3):237 -- 267

\bibitem[{Kochenberger et~al(2014)Kochenberger, Hao, Glover, Lewis, L{\"u},
  Wang, and Wang}]{Kochenberger2014}
Kochenberger G, Hao JK, Glover F, Lewis M, L{\"u} Z, Wang H, Wang Y (2014) The
  unconstrained binary quadratic programming problem: a survey. Journal of
  Combinatorial Optimization 28(1):58--81

\bibitem[{Zhu et~al(2016)Zhu, Fang, and Katzgraber}]{ZCH16}
Zhu Z, Fang C, Katzgraber HG (2016) borealis-a generalized global update
  algorithm for boolean optimization problems. arXiv preprint arXiv:160509399

\bibitem[{Boros and Hammer(2002)}]{EP02}
Boros E, Hammer PL (2002) Pseudo-boolean optimization. Discrete Applied
  Mathematics 123(1--3):155 -- 225

\bibitem[{Bian et~al(2010)Bian, Chudak, Macready, and Rose}]{bian2010ising}
Bian Z, Chudak F, Macready WG, Rose G (2010) The ising model: teaching an old
  problem new tricks. D-Wave Systems 2

\bibitem[{Zalka(1999)}]{Zal99}
Zalka C (1999) Grover's quantum searching algorithm is optimal. Phys Rev A
  60:2746--2751

\bibitem[{Venturelli et~al(2015)Venturelli, Mandr\`a, Knysh, O'Gorman, Biswas,
  and Smelyanskiy}]{VMK+15}
Venturelli D, Mandr\`a S, Knysh S, O'Gorman B, Biswas R, Smelyanskiy V (2015)
  Quantum optimization of fully connected spin glasses. Phys Rev X 5:031,040

\bibitem[{Katzgraber et~al(2015)Katzgraber, Hamze, Zhu, Ochoa, and
  Munoz-Bauza}]{KHZ+15}
Katzgraber HG, Hamze F, Zhu Z, Ochoa AJ, Munoz-Bauza H (2015) Seeking quantum
  speedup through spin glasses: The good, the bad, and the ugly. Phys Rev X
  5:031,026

\bibitem[{Denchev et~al(2016)Denchev, Boixo, Isakov, Ding, Babbush,
  Smelyanskiy, Martinis, and Neven}]{DBI+16}
Denchev VS, Boixo S, Isakov SV, Ding N, Babbush R, Smelyanskiy V, Martinis J,
  Neven H (2016) What is the computational value of finite-range tunneling?
  Phys Rev X 6:031,015

\bibitem[{King et~al(2017)King, Yarkoni, Raymond, Ozfidan, King, Nevisi,
  Hilton, and McGeoch}]{KYRO+17}
King J, Yarkoni S, Raymond J, Ozfidan I, King AD, Nevisi MM, Hilton JP, McGeoch
  CC (2017) Quantum annealing amid local ruggedness and global frustration.
  arXiv preprint arXiv:170104579

\bibitem[{van Dam et~al(2001)van Dam, Mosca, and Vazirani}]{DMV01}
van Dam W, Mosca M, Vazirani U (2001) How powerful is adiabatic quantum
  computation? In: IEEE Symposium on Foundations of Computer Science, 2001.
  Proceedings. 42nd, pp 279--287

\bibitem[{Baritompa et~al(2005)Baritompa, Bulger, and Wood}]{BBW05}
Baritompa WP, Bulger DW, Wood GR (2005) Grover's quantum algorithm applied to
  global optimization. SIAM Journal on Optimization 15(4):1170--1184

\bibitem[{Farhi et~al(2014)Farhi, Goldstone, and Gutmann}]{FGG14}
Farhi E, Goldstone J, Gutmann S (2014) A quantum approximate optimization
  algorithm. arXiv preprint arXiv:14114028

\bibitem[{Temme et~al(2011)Temme, Osborne, Vollbrecht, Poulin, and
  Verstraete}]{TOV+11}
Temme K, Osborne TJ, Vollbrecht KG, Poulin D, Verstraete F (2011) Quantum
  metropolis sampling. Nature 471(7336):87--90

\bibitem[{Cipra(1987)}]{Cip87}
Cipra BA (1987) An introduction to the ising model. Am Math Monthly
  94(10):937--959

\bibitem[{Farhi et~al(2001)Farhi, Goldstone, Gutmann, Lapan, Lundgren, and
  Preda}]{FGG+01}
Farhi E, Goldstone J, Gutmann S, Lapan J, Lundgren A, Preda D (2001) A quantum
  adiabatic evolution algorithm applied to random instances of an np-complete
  problem. Science 292(5516):472--475

\bibitem[{Farhi et~al(2002{\natexlab{a}})Farhi, Goldstone, and
  Gutmann}]{farhi2002quantum}
Farhi E, Goldstone J, Gutmann S (2002{\natexlab{a}}) Quantum adiabatic
  evolution algorithms versus simulated annealing. arXiv preprint
  quant-ph/0201031

\bibitem[{Farhi et~al(2002{\natexlab{b}})Farhi, Goldstone, and
  Gutmann}]{farhi2002quantum2}
Farhi E, Goldstone J, Gutmann S (2002{\natexlab{b}}) Quantum adiabatic
  evolution algorithms with different paths. arXiv preprint quant-ph/0208135

\bibitem[{Trummer and Koch(2016)}]{TK16}
Trummer I, Koch C (2016) Multiple query optimization on the d-wave 2x adiabatic
  quantum computer. Proc VLDB Endow 9(9):648--659

\bibitem[{Trotter(1959)}]{Tro59}
Trotter HF (1959) On the product of semi-groups of operators. Proceedings of
  the American Mathematical Society 10(4):545--551

\bibitem[{Suzuki(1990)}]{Suz90}
Suzuki M (1990) Fractal decomposition of exponential operators with
  applications to many-body theories and monte carlo simulations. Physics
  Letters A 146(6):319 -- 323

\bibitem[{Barahona(1982)}]{Bar82}
Barahona F (1982) On the computational complexity of ising spin glass models.
  Journal of Physics A: Mathematical and General 15(10):3241

\bibitem[{Barends et~al(2016)Barends, Shabani, Lamata, Kelly, Mezzacapo, Heras,
  Babbush, Fowler, Campbell, Chen, Chen, Chiaro, Dunsworth, Jeffrey, Lucero,
  Megrant, Mutus, Neeley, Neill, O'Malley, Quintana, Roushan, Sank,
  Vainsencher, Wenner, White, Solano, Neven, and Martinis}]{BSL+16}
Barends R, Shabani A, Lamata L, Kelly J, Mezzacapo A, Heras UL, Babbush R,
  Fowler AG, Campbell B, Chen Y, Chen Z, Chiaro B, Dunsworth A, Jeffrey E,
  Lucero E, Megrant A, Mutus JY, Neeley M, Neill C, O'Malley PJJ, Quintana C,
  Roushan P, Sank D, Vainsencher A, Wenner J, White TC, Solano E, Neven H,
  Martinis JM (2016) Digitized adiabatic quantum computing with a
  superconducting circuit. Nature 534(7606):222--226

\bibitem[{Berry et~al(2015{\natexlab{a}})Berry, Childs, Cleve, Kothari, and
  Somma}]{BCC+15}
Berry DW, Childs AM, Cleve R, Kothari R, Somma RD (2015{\natexlab{a}})
  Simulating hamiltonian dynamics with a truncated taylor series. Phys Rev Lett
  114:090,502

\bibitem[{Berry et~al(2015{\natexlab{b}})Berry, Childs, and Kothari}]{BCK15}
Berry DW, Childs AM, Kothari R (2015{\natexlab{b}}) Hamiltonian simulation with
  nearly optimal dependence on all parameters. In: Foundations of Computer
  Science (FOCS), 2015 IEEE 56th Annual Symposium on, pp 792--809

\bibitem[{Berry et~al(2014)Berry, Childs, Cleve, Kothari, and Somma}]{BCC+14}
Berry DW, Childs AM, Cleve R, Kothari R, Somma RD (2014) Exponential
  improvement in precision for simulating sparse hamiltonians. In: Proceedings
  of the 46th Annual ACM Symposium on Theory of Computing, ACM, New York, NY,
  USA, STOC '14, pp 283--292

\bibitem[{Boyer et~al(1998)Boyer, Brassard, H{\o}yer, and Tapp}]{BBH+98}
Boyer M, Brassard G, H{\o}yer P, Tapp A (1998) Tight bounds on quantum
  searching. Fortschritte der Physik 46(4-5):493--505

\bibitem[{Liu and Koehler(2010)}]{Liu2010620}
Liu Y, Koehler GJ (2010) Using modifications to grover's search algorithm for
  quantum global optimization. European Journal of Operational Research
  207(2):620 -- 632

\bibitem[{Durr and Hoyer(1996)}]{DH96}
Durr C, Hoyer P (1996) A quantum algorithm for finding the minimum. arXiv
  preprint quant-ph/9607014

\bibitem[{Mermin(2007)}]{Mer07}
Mermin ND (2007) Quantum computer science: an introduction. Cambridge
  University Press

\bibitem[{Strubell(2011)}]{Stu11}
Strubell E (2011) An introduction to quantum algorithms. COS498 Chawathe Spring

\bibitem[{Ambainis(2004)}]{Amb04}
Ambainis A (2004) Quantum search algorithms. SIGACT News 35(2):22--35

\bibitem[{Hogg and Portnov(2000)}]{HP00}
Hogg T, Portnov D (2000) Quantum optimization. Inf Sci Inf Comput Sci
  128(3-4):181--197

\bibitem[{Zahedinejad et~al(2014)Zahedinejad, Schirmer, and Sanders}]{ZSS14}
Zahedinejad E, Schirmer S, Sanders BC (2014) Evolutionary algorithms for hard
  quantum control. Phys Rev A 90:032,310

\bibitem[{Brassard et~al(2002)Brassard, Hoyer, Mosca, and Tapp}]{BHM+02}
Brassard G, Hoyer P, Mosca M, Tapp A (2002) Quantum amplitude amplification and
  estimation. Contemporary Mathematics 305:53--74

\bibitem[{Farhi and Harrow(2016)}]{FH16}
Farhi E, Harrow AW (2016) Quantum supremacy through the quantum approximate
  optimization algorithm. arXiv preprint arXiv:160207674

\bibitem[{Wecker et~al(2016)Wecker, Hastings, and Troyer}]{WHT16}
Wecker D, Hastings MB, Troyer M (2016) Training a quantum optimizer. arXiv
  preprint arXiv:160505370

\bibitem[{Yang et~al(2016)Yang, Rahmani, Shabani, Neven, and Chamon}]{YRS+16}
Yang ZC, Rahmani A, Shabani A, Neven H, Chamon C (2016) Optimizing variational
  quantum algorithms using pontryagin's minimum principle. arXiv preprint
  arXiv:160706473

\bibitem[{Bryson(1975)}]{Bry75}
Bryson AE (1975) Applied optimal control: optimization, estimation and control.
  CRC Press

\bibitem[{Stengel(2012)}]{Ste12}
Stengel RF (2012) Optimal control and estimation. Courier Corporation

\bibitem[{McClean et~al(2016)McClean, Romero, Babbush, and
  Aspuru-Guzik}]{JJR+16}
McClean JR, Romero J, Babbush R, Aspuru-Guzik A (2016) The theory of
  variational hybrid quantum-classical algorithms. New Journal of Physics
  18(2):023,023

\bibitem[{Pontryagin(1987)}]{Pon87}
Pontryagin LS (1987) Mathematical theory of optimal processes. CRC Press

\bibitem[{Palittapongarnpim et~al(2016)Palittapongarnpim, Wittek, Zahedinejad,
  and Sanders}]{PWZ+16}
Palittapongarnpim P, Wittek P, Zahedinejad E, Sanders BC (2016) Learning in
  quantum control: High-dimensional global optimization for noisy quantum
  dynamics. arXiv preprint arXiv:160703428

\bibitem[{Crosson et~al(2014)Crosson, Farhi, Lin, Lin, and Shor}]{CFL+14}
Crosson E, Farhi E, Lin CYY, Lin HH, Shor P (2014) Different strategies for
  optimization using the quantum adiabatic algorithm. arXiv preprint
  arXiv:14017320

\bibitem[{Karimi and Rosenberg(2017)}]{KR16}
Karimi H, Rosenberg G (2017) Boosting quantum annealer performance via sample
  persistence. Quantum Information Processing 16(7):166

\bibitem[{Karimi et~al(2017)Karimi, Rosenberg, and Katzgraber}]{KGK17}
Karimi H, Rosenberg G, Katzgraber HG (2017) Effective optimization using sample
  persistence: A case study on quantum annealers and various monte carlo
  optimization methods. arXiv preprint arXiv:170607826v1

\bibitem[{Crawford et~al(2016)Crawford, Levit, Ghadermarzy, Oberoi, and
  Ronagh}]{CLG+16}
Crawford D, Levit A, Ghadermarzy N, Oberoi JS, Ronagh P (2016) Reinforcement
  learning using quantum boltzmann machines. arXiv preprint arXiv:161205695

\bibitem[{Amin et~al(2016)Amin, Andriyash, Rolfe, Kulchytskyy, and
  Melko}]{AAR+16}
Amin MH, Andriyash E, Rolfe J, Kulchytskyy B, Melko R (2016) Quantum boltzmann
  machine. arXiv preprint arXiv:160102036

\bibitem[{Kitaev(1995)}]{Kit95}
Kitaev AY (1995) Quantum measurements and the abelian stabilizer problem. arXiv
  preprint quant-ph/9511026

\bibitem[{Barenco et~al(1995)Barenco, Bennett, Cleve, DiVincenzo, Margolus,
  Shor, Sleator, Smolin, and Weinfurter}]{BBC+95}
Barenco A, Bennett CH, Cleve R, DiVincenzo DP, Margolus N, Shor P, Sleator T,
  Smolin JA, Weinfurter H (1995) Elementary gates for quantum computation. Phys
  Rev A 52:3457--3467

\bibitem[{Marriott and Watrous(2005)}]{MW05}
Marriott C, Watrous J (2005) Quantum arthur---merlin games. Comput Complex
  14(2):122--152

\bibitem[{Magniez et~al(2007)Magniez, Santha, and Szegedy}]{MSS07}
Magniez F, Santha M, Szegedy M (2007) Quantum algorithms for the triangle
  problem. SIAM Journal on Computing 37(2):413--424

\bibitem[{Gall(2014)}]{Gal14}
Gall FL (2014) Improved quantum algorithm for triangle finding via
  combinatorial arguments. In: Foundations of Computer Science (FOCS), 2014
  IEEE 55th Annual Symposium on, pp 216--225

\bibitem[{Childs and Eisenberg(2003)}]{CE03}
Childs AM, Eisenberg JM (2003) Quantum algorithms for subset finding. arXiv
  preprint quant-ph/0311038

\bibitem[{D{\"u}rr et~al(2004)D{\"u}rr, Heiligman, H{\o}yer, and
  Mhalla}]{DHH+04}
D{\"u}rr C, Heiligman M, H{\o}yer P, Mhalla M (2004) Quantum Query Complexity
  of Some Graph Problems, Springer Berlin Heidelberg, Berlin, Heidelberg, pp
  481--493

\bibitem[{Cerf et~al(2000)Cerf, Grover, and Williams}]{CGW00}
Cerf NJ, Grover LK, Williams CP (2000) Nested quantum search and structured
  problems. Phys Rev A 61:032,303

\bibitem[{D{\"o}rn(2005)}]{Dor05}
D{\"o}rn S (2005) Quantum complexity bounds for independent set problems. arXiv
  preprint quant-ph/0510084

\bibitem[{Lloyd et~al(2013)Lloyd, Mohseni, and Rebentrost}]{LMR13}
Lloyd S, Mohseni M, Rebentrost P (2013) Quantum algorithms for supervised and
  unsupervised machine learning. arXiv preprint qunat-ph/13070411v2

\bibitem[{Sch\"{o}ning(1999)}]{Sch99}
Sch\"{o}ning U (1999) A probabilistic algorithm for k-sat and constraint
  satisfaction problems. In: Proceedings of the 40th Annual Symposium on
  Foundations of Computer Science, IEEE Computer Society, Washington, DC, USA,
  FOCS '99, pp 410--

\bibitem[{Paturi et~al(1997)Paturi, Pudlak, and Zane}]{PPZ97}
Paturi R, Pudlak P, Zane F (1997) Satisfiability coding lemma. In: Foundations
  of Computer Science, 1997. Proceedings., 38th Annual Symposium on, pp
  566--574

\bibitem[{Paturi et~al(1998)Paturi, Pudlik, Saks, and Zane}]{PPS+98}
Paturi R, Pudlik P, Saks ME, Zane F (1998) An improved exponential-time
  algorithm for k-sat. In: Foundations of Computer Science, 1998. Proceedings.
  39th Annual Symposium on, pp 628--637

\bibitem[{Dantsin and Wolpert(2005)}]{Ber05}
Dantsin E, Wolpert A (2005) An Improved Upper Bound for SAT, Springer Berlin
  Heidelberg, Berlin, Heidelberg, pp 400--407

\bibitem[{Dantsin et~al(2005)Dantsin, Kreinovich, and Wolpert}]{DKW05}
Dantsin E, Kreinovich V, Wolpert A (2005) On quantum versions of
  record-breaking algorithms for sat. SIGACT News 36(4):103--108

\bibitem[{Back and Khuri(1994)}]{BK94}
Back T, Khuri S (1994) An evolutionary heuristic for the maximum independent
  set problem. In: Evolutionary Computation, 1994. IEEE World Congress on
  Computational Intelligence., Proceedings of the First IEEE Conference on, pp
  531--535 vol.2

\bibitem[{Johnson(1979)}]{Joh79}
Johnson DS (1979) Computers and intractability: A guide to the theory of
  np-completeness

\bibitem[{Robson(2001)}]{Rob01}
Robson JM (2001) Finding a maximum independent set in time $\mathcal{O}(2n/4)$.
  Tech. rep., Technical Report 1251-01, LaBRI, Universit{\'e} Bordeaux I

\bibitem[{Wittek(2014)}]{Wittek14}
Wittek P (2014) Quantum Machine Learning. Academic Press, Boston

\bibitem[{Bulger(2005)}]{Bul05}
Bulger D (2005) Quantum basin hopping with gradient-based local optimisation.
  arXiv preprint quant-ph/0507193

\bibitem[{Bulger(2007)}]{Bul07}
Bulger DW (2007) Combining a local search and grover's algorithm in black-box
  global optimization. J Optim Theory Appl 133(3):289--301

\end{thebibliography}

\end{document}